\documentclass[prd,twocolumn,amsmath,amssymb,floatfix,superscriptaddress]{revtex4-1}

\usepackage{graphicx}
\usepackage{amssymb}
\usepackage{amsmath}
\usepackage{color}
\usepackage{ wasysym }
\usepackage{hyperref}
\usepackage{tabu}

\def\barray{\begin{array}}
	\def\earray{\end{array}}
\def\be{\begin{equation}}
\def\ee{\end{equation}}
\def\ben{\begin{equation} \nonumber}
\def\een{\end{equation}}
\def\ban{\begin{eqnarray*}}
	\def\ean{\end{eqnarray*}}
\def\ba{\begin{eqnarray}}
\def\ea{\end{eqnarray}}

\def\({\left(}
\def\){\right)}

\def\mnras{MNRAS}

\def\physrep{Phys. Rep.}

\def\apjs{ApJS}

\usepackage{lineno}
\definecolor{valecol}{rgb}{0,0.5, 1.}

\newcommand{\jt}[1]{{\textcolor{black}{#1}}}
\newcommand{\mm}[1]{{\textcolor{black}{#1}}}
\graphicspath{{./fig/}}

\begin{document}
	
	\title{Can high-redshift Hubble diagrams rule out the standard model of cosmology in the context of cosmographic method? }
	
	\author{S. Pourojaghi}
	\affiliation{Department of Physics, Bu-Ali Sina University, Hamedan
		65178, Iran}
	\author{N. F. Zabihi}
	\affiliation{Department of Physics, Bu-Ali Sina University, Hamedan
		65178, Iran}
	\author{M. Malekjani}
	\email{malekjani@basu.ac.ir}
	\affiliation{Department of Physics, Bu-Ali Sina University, Hamedan
		65178, Iran}

\begin{abstract}
Using mock data for the Hubble diagrams of type Ia supernovae (SNIa) and quasars (QSOs) generated based on the standard model of cosmology, and using the least-squares method based on the Markov-Chain-Monte-Carlo (MCMC) algorithm, we first put constraints on the cosmographic parameters in the context of the various model-independent cosmographic methods reconstructed from the Taylor $4^{th}$ and $5^{th}$ order expansions and the Pade (2,2) and (3,2) polynomials of the Hubble parameter, respectively. We then reconstruct the distance modulus in the framework of cosmographic methods and calculate the percentage difference between the distance modulus of the cosmographic methods and that of the standard model. The percentage difference is minimized when the Pade approximation is used which means that the Pade cosmographic method is sufficiently suitable for reconstructing the distance modulus even at high-redshifts. In the next step, using the real observational data for the Hubble diagrams of SNIa, QSOs, gamma-ray-bursts (GRBs), and observations from baryon acoustic oscillations (BAO) in two sets of the low-redshift combination (SNIa+QSOs+GRBs+BAO) embracing the redshift range of $0.01<z<2.26$ and the high-redshift combination (SNIa+QSOs+GRBs) which covers a redshift range of $0.01< z < 5.5$, we put observational constraints on the cosmographic parameters of the Pade cosmography and also the standard model. Our analysis indicates that Pade cosmographic approaches do not reveal any cosmographic tension between the standard model and the observational data. We also confirm this result, using the statistical AIC criteria. Finally, we put the cosmographic method in the redshift-bin data and find a larger value of $\Omega_{m0}$ extracted from $s_0$ parameter compared with those of the $q_0$ parameter and Planck-$\Lambda$CDM values.
	
\end{abstract}
\maketitle

\section{Introduction}

In 1998, observations of SNIa confirmed that our universe is undergoing an accelerated expansion \citep{Riess:1998cb,Perlmutter:1998np,Kowalski:2008ez}. In addition, this accelerated phase has been approved by a large body of the cosmic observations including Wilkinson the Microwave Anisotropy Probe (WMAP) and the Planck observations of the cosmic microwave background (CMB) \citep{Komatsu2009,Jarosik:2010iu,Ade:2015rim}, the Sloan Digital Sky Survey (SDSS), WMAP, 2dFGRS, and 
6dFGS observations of the large-scale structure (LSS), and the baryon acoustic oscillations (BAO) measurements \citep{Tegmark:2003ud,Cole:2005sx,Eisenstein:2005su,Percival2010,Blake:2011en,Reid:2012sw}, high-redshift galaxies \citep{Alcaniz:2003qy}, high-redshift galaxy clusters \citep{Wang:1998gt,Allen2004} and weak gravitational lensing \citep{Benjamin:2007ys,Amendola:2007rr,Fu:2007qq}.\\
In order to interpret the accelerated expansion of the universe, some cosmological groups modified the standard theory of gravity and considered that the theory of general relativity (GR) encounters problems on cosmological scales \citep{Gu:2001ni,Sotiriou:2008rp,Nojiri:2010wj,Clifton:2011jh,DeFelice:2010aj}. For a recent review see \citep{Shankaranarayanan:2022wbx}. In contrast, other groups have considered an unknown substance with negative pressure the so-called dark energy (DE) as a responsible for this acceleration. From the later point of view, the simplest cosmological model is the $\Lambda$CDM model, also known as the concordance model, which has a constant equation of state (EoS) for DE equal to $-1$. In this model, the $\Lambda$ and CDM stand for cosmological constant and cold dark matter, respectively. Even though the $\Lambda$CDM cosmology is in good agreement with a large body of the observational data, it suffers from fundamental problems like fine-tuning and cosmic coincidence \citep{Weinberg:1988cp,Sahni:1999gb,Carroll:2000fy,Padmanabhan2003,Copeland:2006wr}. In addition, recently a conflict in the Hubble constant value has been emerged by CMB measurements of Planck with classical distance ladder \citep{Freedman:2017yms}. These discrepancies have led to cosmological tensions, which have been a popular topic lately, and a wide variety of DE models with time-varying EoS have been emerged so far to solve these problems {\citep{Veneziano:1979ec, Erickson:2001bq, Armendariz-Picon:2000ulo, Thomas:2002pq, Caldwell:1999ew, Padmanabhan:2002cp, Gasperini:2001pc, Elizalde:2004mq, Gomez-Valent:2014fda}. For a recent review about the cosmological tensions of the standard model of cosmology, we refer the reader to \citep{Abdalla:2022yfr}. In recent years, many efforts have been made to obtain the best cosmological model with the lowest inconsistency with observational data. By comparing different models, some have been ruled out, though some showed good agreement with the observational data (see  {\citep{Malekjani:2016edh, Rezaei:2017yyj, Malekjani:2018qcz, Rezaei:2019roe, Lin:2019htv, Rezaei:2019xwo}}).\\
While the expansion of the universe can be investigated model dependently, one can study the expansion history of the universe directly from observational data without considering any particular model. These approaches are recognized as model-independent methods.\\
Gaussian process is a non-linear Bayesian approach that can directly reconstruct observational data. The method is based on the distribution over function, and a covariance function will be included to connect two distinct data points, and the process will progress to predict other data values in higher redshifts {\citep{Zhang:2018gjb, Rasmussen:2006, Seikel:2012uu}}. This approach has been applied broadly in cosmology {\citep{Seikel:2012uu, Shafieloo:2012ht, CMS:2018skt, Liao:2019qoc, Mehrabi:2020zau, Mehrabi:2021cob}}.\\
Another model-independent method is the smoothing method, a non-parametric iterative approach that reconstructs functions from an initial ansatz. It uses a smoothing kernel, and each step results in a better fit for the cosmological data \citep{Shafieloo:2007cs}. Genetic algorithm is another model-independent approach inspired by natural selection used in \citep{Nesseris:2010ep} on supernova Ia data to reconstruct the Hubble parameter H(z). More details and applications of the Genetic algorithm in cosmology can be found in \cite{Bogdanos:2009ib}. Artificial neural network (ANN) can be considered as another model-independent method that has also been used as a non-parametric reconstruction of the cosmological functions that does not assume any statistical distribution of observational data \cite{Gomez-Vargas:2021zyl}.\\
 Finally, the cosmographic approach, which would be used in this paper as a model-independent approach, is basically defined on the basis of the Taylor expansion of cosmological functions of scale factor $a$ (or equivalently cosmic redshift $z$) such as Hubble parameter and luminosity distance around the present time $z=0$. This approach has commonly been used in literature \citep{Sahni:2002fz, Alam:2003sc,Cattoen:2007sk, Capozziello:2009ka, Capozziello:2011tj, Capozziello:2018jya, Benetti:2019gmo, Escamilla-Rivera:2019aol, Lusso:2019akb, Rezaei:2020lfy, Bargiacchi:2021fow, Capozziello:2021xjw}. This method relies only on the assumption of a homogeneous and isotropic universe described in the Friedman-Lemaitre-Robertson-Walker (FLRW) metric. While a Taylor expansion with respect to redshift z can estimate the cosmic evolution at $z\sim 0$, it fails at high redshifts (see \cite{Caldwell:2004vi,SupernovaSearchTeam:2004lze,Visser:2003vq} for earlier attempts). The main issue of Taylor series in $z$-redshift, is the convergence problem at high redshifts causes to serve the error propagation in Taylor series and consequently reduces the cosmography prediction \cite{Cattoen:2007sk,Busti:2015xqa,Capozziello:2019cav,Li:2019qic}. One possible way to improve the cosmography method is the use of the auxiliary variables to re-parameterize the redshift variable through functions of z, for example y-redshift as $y=z/(1+z)$ \cite{Capozziello:2011tj}. The other way is to assume a smooth evolution of the observable quantities by expanding them in terms of rational approximations like Pade \cite{Gruber:2013wua,Wei:2013jya,Capozziello:2019cav,Capozziello:2018jya,Benetti:2019gmo,Capozziello:2020ctn} and Chebyshev polynomials \cite{Capozziello:2017nbu}.
%\textcolor{green}{The authors of \citep{Sahni:2002fz} and \citep{Alam:2003sc} introduced this method to discriminate between dark energy models for the first time. Capozziello and Salzano \citep{Capozziello:2009ka} examined the cosmographic approach as a valuable tool to differentiate between General Relativity and alternative theories. By assessing $f(R)$ gravity, they showed that the cosmographic parameters could constrain cosmological models. Furthermore, the authors of \citep{Capozziello:2011tj} studied the expansion scenario of the universe using the observations from SNIa, $H(z)$ data from differential galaxy ages, GRBs and BAO and showed that the cosmographic method is trustworthy for extracting cosmological information. They also reported that there might be a possibility of deviation from the $\Lambda$CDM model \citep{Capozziello:2011tj}. }

 Recently, Lusso, et. al., \citep{Lusso:2019akb} by using the high-redshift Hubble diagrams of QSOs and GRBs claimed \jt{the existence of} a big tension between the cosmographic parameters of the standard $\Lambda$CDM model and those of the cosmographic method defined on the basis of the logarithmic expansion of the luminosity distance in terms of y-redshift \citep[see also][]{Risaliti:2018reu}. \mm{ Notice that the result of \cite{Lusso:2019akb} is impacted by cosmographic method, as explained in \cite{Yang:2019vgk}.} The cosmographic approach presented in \citep{Lusso:2019akb}, has been extended to a general case by
applying the orthogonalized logarithmic polynomials of the luminosity distance \citep{Bargiacchi:2021fow}. In this general case, the authors of \citep{Bargiacchi:2021fow} showed a big tension ($>4\sigma$) between the standard model and model-independent cosmographic method using the Hubble diagrams of SNIa and QSOs. \mm{ It should be emphasized that the above tension depends on the slope parameter of the log-linear relation between the UV and X-ray luminosities of QSOs described by $\log{L_X}=\gamma \log{L_{UV} + \beta}$. The non-evolution treatment of this relation against cosmic redshift has been tested in \cite{Risaliti:2018reu}. They drove an average value for the slope parameter $\gamma \sim 0.6$, insensitively from the specific choice of the redshift bins of their analysis. However, it is mentionable that the authors of \cite{Hu:2022udt} showed a slight change of the slope parameter can make the $4\sigma$ tension disappear. In the other word, they found that the QSOs relation can affect the cosmographic constraints}. Going beyond the standard model, Rezaei et al., in \citep{Rezaei:2020lfy} investigated three different DE parameterizations namely wCDM, CPL and Pade parameterization for EoS parameter using the Hubble diagrams of SNIa, QSOs and GRBs in the cosmographic method . Based on the $4^{\it th}$ order Taylor expansion of the Hubble parameter, they showed that these parameterizations are consistent with model-independent cosmographic approach. The same study has been done for the holographic dark energy (HDE) model with time-varying EoS parameter in \citep{Pourojaghi:2021}. In the context of cosmographic method based on the $4^{\it th}$ order Taylor expansion of the Hubble parameter, Pourojaghi and Malekjani showed that there is no tension between HDE model and high-redshift Hubble diagrams of QSOs and GRBs \citep{Pourojaghi:2021}.\\

On the other hand, the authors in \Citep{Banerjee:2020bjq} \jt{ have shown that the logarithmic polynomial expansion of the luminosity distance, which is the basis for the strong claim \Citep{Risaliti:2018reu,Lusso:2019akb,Bargiacchi:2021fow} that there is a deviation of about $4\sigma$ from the flat-$\Lambda$CDM model when fitted to high-redshift Hubble diagrams of QSOs and GRBs, holds up to redshifts $z\sim 2$. In other words, the log polynomial approximation cannot recover the flat-$\Lambda$CDM model at higher redshifts and thus undermines the $4\sigma$ tension claim \cite[see also][]{Yang:2019vgk}. Overall, we should be very careful when applying the cosmographic method at high-redshifts, implying that we should be able to recover the flat-$\Lambda$CDM model.} \jt{It is worth noting that the direct fit of the standard model to the QSOs data shows a deviation of about $4\sigma$ in the best-fit  matter density $\Omega_{m0}=0.9$ from the flat-$\Lambda$CDM value $\Omega_{m0}=0.3$ \cite{Yang:2019vgk}. In this fit, a universe without DE ($\Omega_m=1$) lies within the $1\sigma$ confidence region.} We note that the dynamical DE models are in complete agreement with high-redshift Hubble diagrams of QSOs and GRBs in the context of the cosmographic method \cite{Rezaei:2020lfy,Pourojaghi:2021}. On the basis of the above description, \jt{we should validate the cosmographic approach before getting any conclusion at high-redshifts. In this regard, the authors of \citep{Capozziello:2018jya} showed that the cosmographic approach based on the rational Pade series performs better than the standard Taylor series at redshifts $z>1$.}
In addition to Pade polynomials, the other rational polynomials that have been used in cosmographic method is the Chebyshev approximations \citep{Capozziello:2017nbu}. It has been shown that the rational Chebyshev series works better than standard Taylor series and reduces the error truncation \citep{Capozziello:2017nbu}. In the context of inverse cosmography \citep[see][]{Escamilla-Rivera:2019aol}, the Chebyshev-like EoS parameter of DE mimics the Pade EoS parameter at high-redshifts, but it has a divergence at low-redshifts \citep{Munoz:2020gok}.\\

One possible way to examine the validity of the cosmographic method at high-redshifts is the use of mock data for the Hubble diagram generated upon the cosmological model. In this concern, we expect that the distance modulus of both cosmographic method and cosmological model are to agree with the mock data. In fact any tension between cosmographic method and mock data has not physical meaning and therefore is related to the error truncation of the mathematical approximation used in cosmographic approach. We investigate both Taylor and Pade approximations of the Hubble parameter used in cosmographic method for reconstructing the distance modulus of SNIa, QSOs and GRBs. Using mock data for the Hubble diagrams of SNIa and QSOs produced based on the normal distribution around the mean value $\mu_{\Lambda}(z_i)$, where $\mu_{\Lambda}(z_i)$ is the distance modulus of the standard flat-$\Lambda$CDM model, we can choose the best approximation of Hubble parameter for cosmographic method. The goal of our analysis based on mock data is to show that the rational pade approximation performs better than Taylor series beyond $z=1$, where the standard cosmographic approach suffers from convergence problem. We will show that unlike to Taylor series, the approximation is under control in the rational Pade polynomials.
After fixing the cosmographic method that performs at higher redshift with a minimum error truncation, we compare the best-fit values of the cosmographic parameters of standard $\Lambda$CDM model obtained using the real observational data for the Hubble diagrams of SNIa, QSOs and GRBs, with those of the model-independent cosmographic method constrained with the same data sets. Comparing these two measurements can lead us to get a reasonable conclusion about the consistency of the standard $\Lambda$CDM model with the observational Hubble diagrams from SNIa, QSOs and GRBs.\\
 
The layout of this article is as follows:
In section \ref{sect:cosmography}, we will introduce cosmographic approaches based on the linear Taylor series, and rational Pade series. We then present the cosmographic parameters for cosmographic approaches and also for the standard $\Lambda$CDM model. In section \ref{sect:Mock_Data}, the process of producing mock data for the Hubble diagrams of SNIa and QSOs has been presented. We then study the validation of the cosmographic methods using mock data. In section \ref{sect:Real_Data}, using the combinations of low-redshift and high-redshift observational data, we put observational constraints on the cosmographic parameters of the flat-$\Lambda$CDM model and compare the results with the confidence regions of the model-independent cosmographic method obtained from the same data sets. Ultimately, in section \ref{sect:conlusion}, we conclude this work.\\

\section{The Cosmographic Approach} \label{sect:cosmography}
Recently, many attempts have been made to investigate the model-independent approaches, extracting information directly from the observational data instead of utilizing DE models. Cosmography is one of the model-independent methods, which profits from an uncomplicated assumption of homogeneity and isotropy, and can assist us to investigate the evolution of the universe. By using the Friedman- Lemaitre- Robertson- Walker (FLRW) metric, one can express cosmographic parameters by scale factor derivatives with respect to cosmic time as follows \citep{Visser:2003vq}:

\begin{align}\label{eq1}
	H = \frac{a^{(1)}}{a}\;, q = - \frac{a^{(2)}}{aH^2}\;, j = \frac{a^{(3)}}{aH^3}\;,\nonumber\\
	s = \frac{a^{(4)}}{aH^4}\;,	l = \frac{a^{(5)}}{aH^5}\;, m = \frac{a^{(6)}}{aH^6}\;,
\end{align}

Where $a^{(n)}$ denotes the $n^{th}$ derivatives of the scale factor. We note that cosmography parameters are independent of dark energy and its equation of state. Hence the cosmography procedure is purely independent of the model. The cosmographic parameters pointed above are extremely valuable observables for extracting information from the universe when calculated at the present time. Moreover, each has a physical meaning behind them, making them proper for explaining the expansion history of the universe. Hubble function with the equation of $H=\frac{\dot{a}}{a}$ indicates the  expansion or contraction phase of the universe ($\dot{a}>0 \rightarrow H>0 \rightarrow$ expansion).  Sign of the deceleration parameter, $q = -\frac{\ddot{a}}{aH^{2}}$, controls the accelerated or decelerated expansion phase of the universe. When $q<0$, $\ddot{a}>0$ reveals that the universe is in accelerating phase. Other cosmographic parameters will become important at higher redshifts. For more information about the physical meanings of cosmographic parameters, we refer the reader to \citep{Pourojaghi:2021}.\\
\subsection{Taylor series for cosmography}
The Taylor expansion of the scale factor up to sixth order in terms of cosmographic parameters takes the form below:

\begin{eqnarray}\label{eq2}
	a(t) \simeq 1+H_0(t-t_0) - \frac{q_0}{2!}H_0^2(t-t_0)^2 + \nonumber\\ \frac{j_0}{3!}H_0^3(t-t_0)^3 +
	\frac{s_0}{4!}H_0^4(t-t_0)^4 +\nonumber\\ \frac{l_0}{5!}H_0^5(t-t_0)^5 +  \frac{m_0}{6!}H_0^6(t-t_0)^6\;.
\end{eqnarray}

We can obtain the following relations between various time-derivative of Hubble parameter $H$ and cosmographic parameters

\begin{eqnarray}\label{eq3}
	&&H^{(1)}=-H^2(q + 1) \nonumber \\
	&&H^{(2)}=H^3(j + 3 q + 2) \nonumber \\
	&&H^{(3)}=H^4(-4j-3q^{2}-12q+s-6) \nonumber \\
	&&H^{(4)}=H^5(10jq+20j+l+30q^{2}+60q-5s+24)\nonumber\\
	&&H^{(5)}=H^6(-10j^{2}-120jq-120j-6l+m-30q^{3}-270\nonumber \\ 
	&&q^{2}+15qs-360q+ 30 s-120)\;.
\end{eqnarray}
While $H^{(n)}$ indicates the $n_{th}$ time derivative of the Hubble parameter $H$. In a cosmographic method, we  reconstruct the Hubble expansion of the universe by Taylor expanding of the Hubble parameter around the present time ($z=0$) as
\begin{eqnarray}\label{eq4}
&&	H(z) \simeq H_0 + H_{1}\vert_{z=0}z + H_{2}\vert_{z=0}\frac{z^2}{2!} +\nonumber\\
&&	 H_{3}\vert_{z=0}\frac{z^3}{3!} + H_{4}\vert_{z=0} \frac{z^4}{4!} +\nonumber\\
&&	  H_{5}\vert_{z=0} \frac{z^5}{5!}\;,
\end{eqnarray}
where $H_{n}$ can be written as follows:
\begin{eqnarray}\label{eq5}
	&&H_1\vert_{z=0}=\frac{dH}{dz}\vert_{z=0}= H_{0}(1+q_0)\nonumber\\
	&&H_2\vert_{z=0}=\frac{d^{2}H}{dz^2}\vert_{z=0}=H_{0} (j_0-q_0^{2})\nonumber\\
	&&H_3\vert_{z=0}=\frac{d^{3}H}{dz^3}\vert_{z=0}=H_{0} (-4j_0q_0-3j_0+3q_0^{3} + 3q_0^{2} -s_0)\nonumber\\
	&&H_4\vert_{z=0}=\frac{d^{4}H}{dz^4}\vert_{z=0}= H_{0} (-4 j_0^{2} + 25 j_0 q_0^{2}+32j_0q_0 +12j_0\nonumber\\
	&&+l_0- 15 q_0^{4}-24q_0^{3}-12q_0^{2}+7q_0s_0+8s_0)\nonumber\\
	&&H_5\vert_{z=0}=\frac{d^{5}H}{dz^5}\vert_{z=0}=H_{0} (70j_0^{2}q_0+60j_0^{2}-210j_0q_0^{3}\nonumber\\
	&&-375j_0q_0^{2}-240j_0q_0+15j_0s_0-60j_0-11l_0q_0-15l_0 \nonumber\\
	&&-m_0+105q_0^{5}+225 q_0^{4} + 180 q_0^{3} - 60 q_0^{2} s_0 + 60 q_0^{2} \nonumber\\
	&& - 105 q_0 s_0 - 60 s_0)\;.
\end{eqnarray}
Because of the divergence problem, we cannot use this expansion at $z>1$, while most of the cosmological data are observed at redshifts higher than $z=1$ \citep{Cattoen:2007sk}. By considering the following y-redshift definition, radius convergence has been improved, while physical definition has not changed \citep{Cattoen:2007sk,Li:2019qic, Capozziello:2020ctn, Vitagliano_2010,Capozziello:2011tj,Rezaei:2020lfy}:
\begin{align}\label{eq6}
	 y=\frac{z}{1+z}\;.
\end{align}
Utilizing Eq.\ref{eq6}, the Taylor expansion of Hubble parameter  around present time ($y=0$) can be rewritten as below:
\begin{eqnarray}\label{eq7}
	&&H(y) \simeq H_0 + \frac{dH}{dy}\vert_{y=0}y + \frac{d^{2}H}{dy^2}\vert_{y=0}\frac{y^2}{2!} + \nonumber\\
	&&\frac{d^{3}H}{dy^3}\vert_{y=0}\frac{y^3}{3!} +\frac{d^{4}H}{dy^4}\vert_{y=0} \frac{y^4}{4!} + \frac{d^{5}H}{dy^5}\vert_{y=0} \frac{y^5}{5!}\;.
\end{eqnarray}
By using Eq. \ref{eq5} and changing derivatives with respect to $z$ into derivatives with respect to $y$, we can reconstruct the Hubble parameter as below:
\begin{eqnarray}\label{eq8}
	E(y) = 1 + C_1y + C_2\frac{y^2}{2!} + C_3\frac{y^3}{3!} + C_4 \frac{y^4}{4!} + C_5 \frac{y^5}{5!}\;,
\end{eqnarray}
where coefficients $C_i$ are given as follows:
\begin{eqnarray}\label{eq9}
	&&C_1= q_{0} + 1\nonumber\\
	&&C_2= j_{0} - q_{0}^{2} + 2 q_{0} + 2\;,\nonumber\\
	&&C_3= - 4 j_{0} q_{0} + 3 j_{0} + 3 q_{0}^{3} - 3 q_{0}^{2} + 6 q_{0} - s_{0} + 6\nonumber\\
	&&C_4= - 4 j_{0}^{2} + 25 j_{0} q_{0}^{2} - 16 j_{0} q_{0} + 12 j_{0} + l_{0} - 15 q_{0}^{4} + 12 q_{0}^{3}\nonumber\\
	&&- 12 q_{0}^{2} + 7 q_{0} s_{0} + 24 q_{0} - 4 s_{0} + 24\nonumber\\
	&&C_5= 70 j_{0}^{2} q_{0} - 20 j_{0}^{2} - 210 j_{0} q_{0}^{3} + 125 j_{0} q_{0}^{2} - 80 j_{0} q_{0} \nonumber\\
	&&+ 15 j_{0} s_{0} + 60 j_{0} - 11 l_{0} q_{0} + 5 l_{0} - m_{0}+ 105 q_{0}^{5} - 75 q_{0}^{4} \nonumber\\
	&&+ 60 q_{0}^{3} - 60 q_{0}^{2} s_{0} - 60 q_{0}^{2} + 35 q_{0} s_{0} + 120 q_{0} - 20 s_{0}\nonumber\\
	&&+ 120\;.
\end{eqnarray}
In section \ref{sect:Mock_Data}, we investigate the $4^{\it th}$ and $5^{\it th}$ order Taylor series based on the $y$ variable and calculate the accuracy of these expansions in terms of cosmic redshift.

\subsection{Rational Pade polynomials for cosmography}
As mentioned before, there are limitations in using cosmography based on Taylor's expansion of z-redshift at high-redshifts. In other words, Taylor's expansion of the Hubble diagram has divergence problems at $z>1$; consequently, the predictions will be unproductive. On the other hand, though utilizing y-redshift in Taylor's expansion improve the high-redshifts divergence problem. Since we cannot continue the Taylor expansion (both z-redshift and y-redshift) to infinity, the error truncation in a particular order causes an inaccuracy in our computations and consequently affects the results. One of the approaches proposed to overcome this limitation is utilizing the rational Pade approximation to reconstruct the Hubble parameter \Citep{Capozziello:2020ctn, Capozziello:2018jya, Benetti:2019gmo, Capozziello:2019cav}.\\
Compared to Taylor's expansion, the Pade parametrization has better efficiency. Since the rational Pade approximation makes divergence amplitude larger, it can increase the convergence domain of the approximation. Thus utilizing it leads to more accurate results in such a way that in cases where Taylor expansion loses efficiency because of divergency, Pade approximation can be a better solution. Typically, a Pade approximation in (n,m) order can be defined as follows:
\begin{eqnarray}\label{eq10}
	P_{n,m}(z)=& \frac{P_0 + P_1 z + ... + P_n z^n}{1 + Q_1 z + ... + Q_m z^m} = \frac{\displaystyle\sum_{i=0}^{n} P_i z^i}{1 + \displaystyle\sum_{j=0}^{m} Q_j z^j}\;,
\end{eqnarray}
where $P$ and $Q$ coefficients correspond to the Taylor expansion coefficients as below:
\begin{align}\label{eq11}
	&P_{n,m}(0) = f(0) \nonumber\\
	&P^{\prime}_{n,m}(0) = f^{\prime}(0) \nonumber\\
	&. \nonumber\\
	&. \nonumber\\
	&. \nonumber\\
	&P^{(n+m)}_{n,m}(0) = f^{(n+m)}(0)\;.
\end{align}
Considering a Taylor expansion as $f(z)=\Sigma_{0}^{\infty} C_{i}z^{i}$ and equalizing it with the Pade approximation, P and Q coefficients can be achieved according to the Taylor series coefficients as follow:
\begin{align}\label{eq12}
	\displaystyle\sum_{i=0}^{\infty} C_i z^i = \frac{\displaystyle\sum_{i=0}^{n} P_i z^i}{1 + \displaystyle\sum_{j=0}^{m} Q_j z^j}\;.
\end{align}
%Also can be written as:
%\begin{align}\label{eq13}
%	(1 + Q_1 z + ... + Q_m z^m)(C_0 + C_1 z + ...)=P_0 + P_1 z + ... + P_n z^n \;.
%\end{align}
Note that in mentioned equalization, we are only allowed to equalize the Pade approximation of order (n,m) with the Taylor series of order $n+m$. Therefore coefficients of Pade approximation based on the Taylor expansion coefficients can be achieved having $n+m+1$ equations and $n+m+1$ unknowns.\\
Ultimately, the dimensionless Hubble parameter can be reconstructed as follows by employing the Pade parametrization:
\begin{align}\label{eq14}
	E(z)= \frac{H}{H_0} = \frac{\displaystyle\sum_{i=0}^{n} P_i z^i}{1 + \displaystyle\sum_{j=0}^{m} Q_j z^j} \;.
\end{align}
 In the following, we consider Pade series (2,2) and (3,2), equivalent to $4^{\it th}$ and $5^{\it th}$ order Taylor series respectively,

\begin{align}
P_{2,2}=& \frac{P_0 + P_1 z + P_2 z^2}{1 + Q_1 z + Q_2 z^2}\;,\nonumber\\
P_0=& 1\;,\nonumber\\
P_1=& E_{1,0} + Q_1\;,\nonumber\\
P_2=& E_{1,0}Q_1 + \frac{E_{2,0}}{2} + Q_2\;,\nonumber\\
Q_1=& \frac{-E_{1,0}E_{1,0} + 2E_{2,0}E_{3,0}}{4E_{1,0}E_{3,0} - 6E_{2,0}^2}\;,\nonumber\\
Q_2=& \frac{3E_{2,0}E_{4,0} - 4E_{3,0}^2}{24E_{1,0}E_{3,0} - 36E_{2,0}^2}\;,
\end{align}

\begin{align}
P_{3,2}=& \frac{P_0 + P_1 z + P_2 z^2 + P_3 z^3}{1 + Q_1 z + Q_2 z^2}\;,\nonumber\\
P_0=& 1\;,\nonumber\\
P_1=& E_{1,0} + Q_1\;,\nonumber\\
P_2=&  E_{1,0}Q_1 + \frac{E_{2,0}}{2} + Q_2\;,\nonumber\\
P_3=& E_{1,0}Q_2 + \frac{E_{2,0}Q_1}{2} + \frac{E_{3,0}}{6}\;,\nonumber\\
Q_1=& \frac{-3E_{2,0}E_{5,0} + 5E_{3,0}E_{4,0}}{15E_{2,0}E_{4,0} - 20E_{3,0}^2}\;,\nonumber\\
Q_2=& \frac{4E_{3,0}E_{5,0} - 5E_{4,0}^2}{60E_{2,0}E_{4,0} - 80E_{3,0}^2}\;,
\end{align}
where $E_{n,0}$ represents $\frac{H_{n}\vert_{z=0}}{H_{0}}$ and $H_{n}\vert_{z=0}$ is calculated from \ref{eq5}.

\subsection{Cosmographic Parameters of the $\Lambda$CDM Model}
So far, we have outlined the cosmographic approach and derived its parameters model-independently. Nevertheless, in order to make a comparison between the cosmographic method and model, we additionally have to extract the cosmographic parameters of model. In the context of GR, by assuming a flat FLRM metric, the Hubble parameter for a universe composed by radiation, matter and DE
can be written as follows:

\begin{eqnarray}\label{eq15}
	&&H^{2}(z)= H^{2}_{0}[\Omega_{m_0}(1+z)^{3}+\;, \nonumber \\
	&&\Omega_{r_0}(1+z)^{4}+(1-\Omega_{m_0}-\Omega_{r_0})e^{3\int{\frac{dz}{1+z}(1+w)}}]\;,
\end{eqnarray}

where $H_{0}$, $\Omega_{m_0}$, and $\Omega_{r_0}$ represent values of the Hubble parameter, pressure-less matter, and radiation energy density, respectively, at the current time. Since we are studying late time evolution, the radiation component portion is negligible compared to other components. The above equation for the flat-$\Lambda$CDM cosmology takes the simple form

\begin{eqnarray}\label{eq16}
	E(z) = \frac{H(z)}{H_0}=\sqrt{\Omega_{m_0}(1+z)^{3}+(1-\Omega_{m_0})}\;.
\end{eqnarray}
Considering derivatives of eq.\ref{eq16}, we can derive the cosmographic parameters for the flat-$\Lambda$CDM model as follow \citep[see also][]{Rezaei:2020lfy,Lusso:2019akb}:

\begin{eqnarray}\label{eq17}
&& q_{0}=-1+\frac{3}{2} \Omega_{m0}\;,\nonumber\\
&&	j_{0}=1\;,\nonumber\\
&&	s_{0}=1-\frac{9}{2} \Omega_{m0}\;,\nonumber\\
&&	l_{0}=1+3\Omega_{m0}+\frac{27}{2} \Omega_{m0}^{2}\;,\nonumber\\
&&	m_0 = - \frac{81}{4} \Omega_{m0}^{3} - 81 \Omega_{m0}^{2} - \frac{27 \Omega_{m0}}{2} + 1\;.	
\end{eqnarray}

\section{Cosmographic method against mock data} \label{sect:Mock_Data}
In this section, we first present the procedure of producing mock data for the Hubble diagrams of SNIa and QSOs based on the flat-$\Lambda$CDM model, and then check the validation of different cosmographic methods considered in this work using mock data. We use the Pantheon sample for supernovae type Ia dataset with redshift ranging $0.01<z<2.26$ that contains 1048 type Ia supernovae gathered from different surveys like Pan-STARRS1, SDSS, SNLS, various low-z, and HST samples \citep{Scolnic:2017caz}. We also use 1598 data points ranging $0.04 < z < 5.1$  for QSOs as the distance indicator from \cite{Risaliti:2018reu}. 
We use the observed redshift and error bars of SNIa and QSOs from the above dataset for generating mock data.\\
 The theoretical distance modulus for a given redshift is as follows:
\begin{eqnarray}\label{eq22}
	\mu (z)= 5 \log_{10}(1+z)\int_{0}^{z} \frac{dz}{E(z)}+\mu_{0}\;,
\end{eqnarray}
where $\mu_{0}= 42.384 - 5 \log_{10}(h)$. We first calculate the distance modulus for the flat-$\Lambda$CDM model, $\mu_{\Lambda}(z_{i})$, in any given redshift $z_{i}$ from Eq.(\ref{eq22}) and by implementing Eq.(\ref{eq16}). Here, we set 
the canonical values $\Omega_{m0}=0.3$ and $h=0.7$. Mock data are generated using a normal distribution with $\mu_{\Lambda}(z_{i})$ as mean and $\Delta\mu(z_{i})$ as standard deviation. Here, $\Delta\mu(z_{i})$ are considered as the observational errors of SNIa and QSOs data points, and $z_{i}$ are their observational redshifts.\\
To ensure that our mock data has been well produced upon the $\Lambda$CDM model, we put constraints on the $\Omega_{m0}$ parameter using mock data and compare the result with the canonical value $\Omega_{m0}=0.3$. To do this, we adopt the standard minimization of chi-square function based on the statistical MCMC algorithm as below:

\begin{eqnarray}\label{eq24}
	\chi^{2}= \sum_{i}\frac{[\mu_{th}(z_{i})-\mu_{moc}(z_{i})]^{2}}{\sigma_{i}^{2}}\;.
\end{eqnarray}

Where $\mu_{th}$ and $\mu_{moc}$ denote theoretical and mock distance modulus, respectively, and $\sigma_{i}$ indicates corresponding errors of mock data. We can confirm the procedure of generating mock data for the Hubble diagrams of SNIa and QSOs, if the canonical value $\Omega_{m0}=0.3$ is in full consistency with $1\sigma$ confidence region of $\Omega_{m0}$ obtained from our MCMC analysis. After confirming that mock data is performing precisely, free parameters of the Taylor and Pade expansions will be constrained using this mock data within the context of MCMC algorithm. Ultimately, distance modulus of standard $\Lambda$CDM and each of the Pade and Taylor expansions can be calculated using the best fit values of the free parameters. The percentage difference between distance modulus of $\Lambda$CDM and each of the expansions which is given by

\begin{eqnarray}\label{eq25}
	\Delta = \frac{\mu_{cosmography}-\mu_{\Lambda}}{\mu_{\Lambda}} \times 100 \;,
\end{eqnarray}
shows the accuracy of the cosmographic method. Lower $\Delta$ for each pair of comparison between expansions and the $\Lambda$CDM model can reveal the minor error truncation, which indicates that the mentioned expansion will work better when used in the model-independent cosmographic approach. In subsequent, we present the numerical results of our analysis.
\begin{figure} 
	\centering
	\includegraphics[width=7.5 cm]{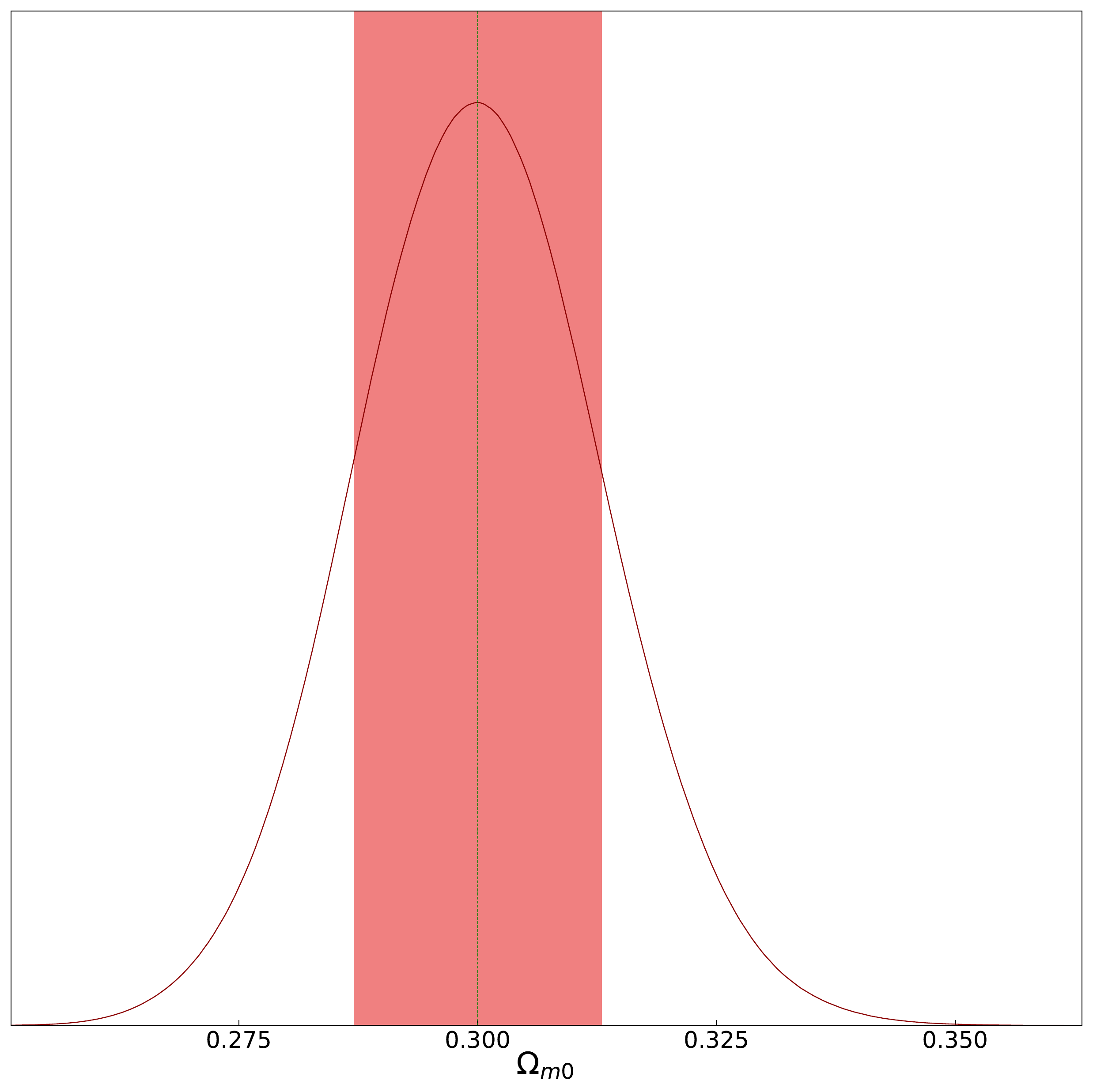}
	\caption{Best-fit value of $\Omega_{m0}$ parameter for the flat-$\Lambda$CDM model using mock SNIa data.}
	\label{fig:pantheon_LCDM}
\end{figure}

\begin{figure} 
	\centering
	\includegraphics[width=7.5 cm]{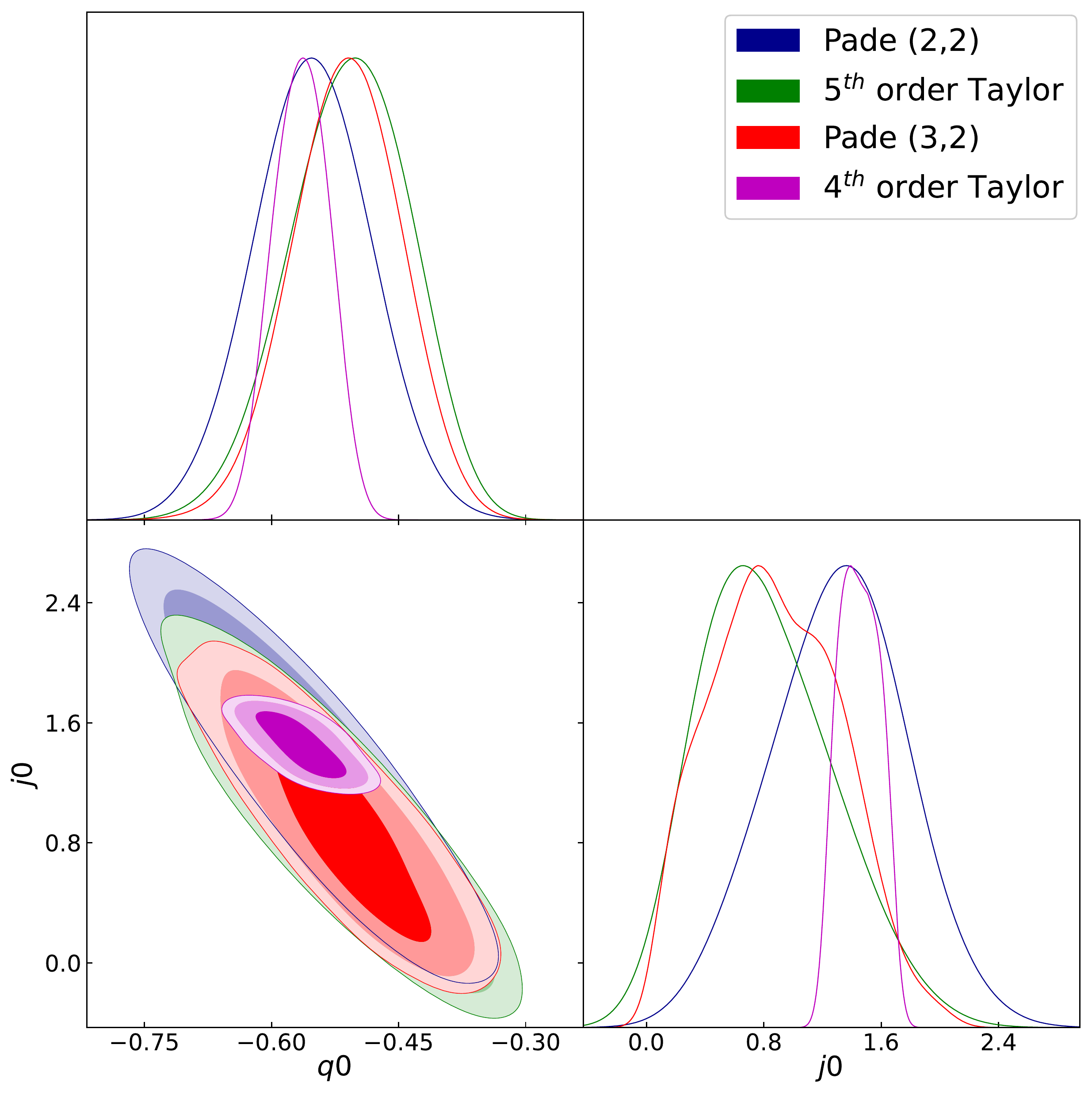}
	\caption{The confidence regions of cosmographic parameters $q_{0}$ and $j_{0}$ obtained form various cosmographic methods using mock SNIa data.}
	\label{fig:fig2}
\end{figure}

\begin{figure*} 
	\centering
	\includegraphics[width=7.6 cm]{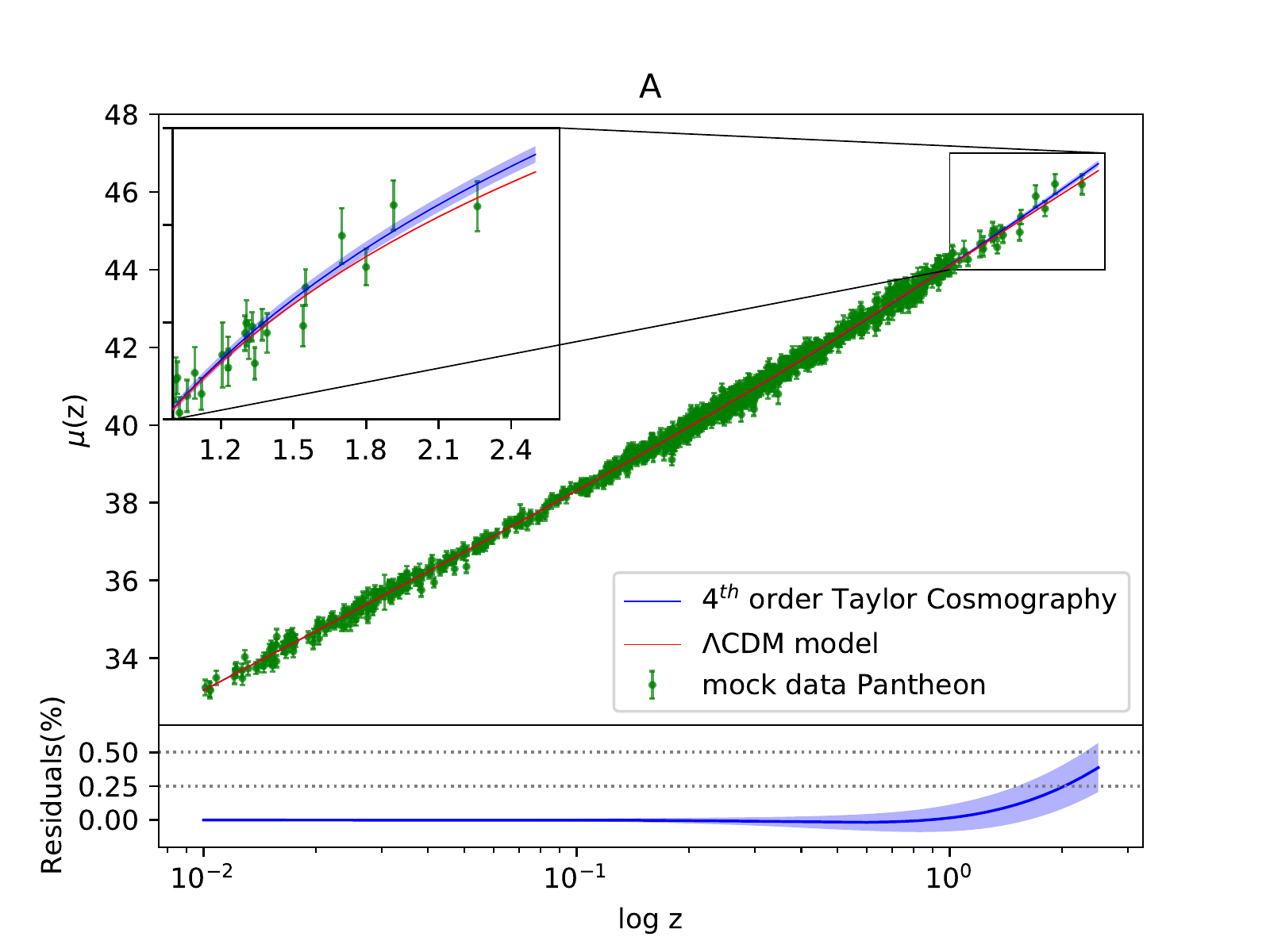}
	\includegraphics[width=6.8 cm]{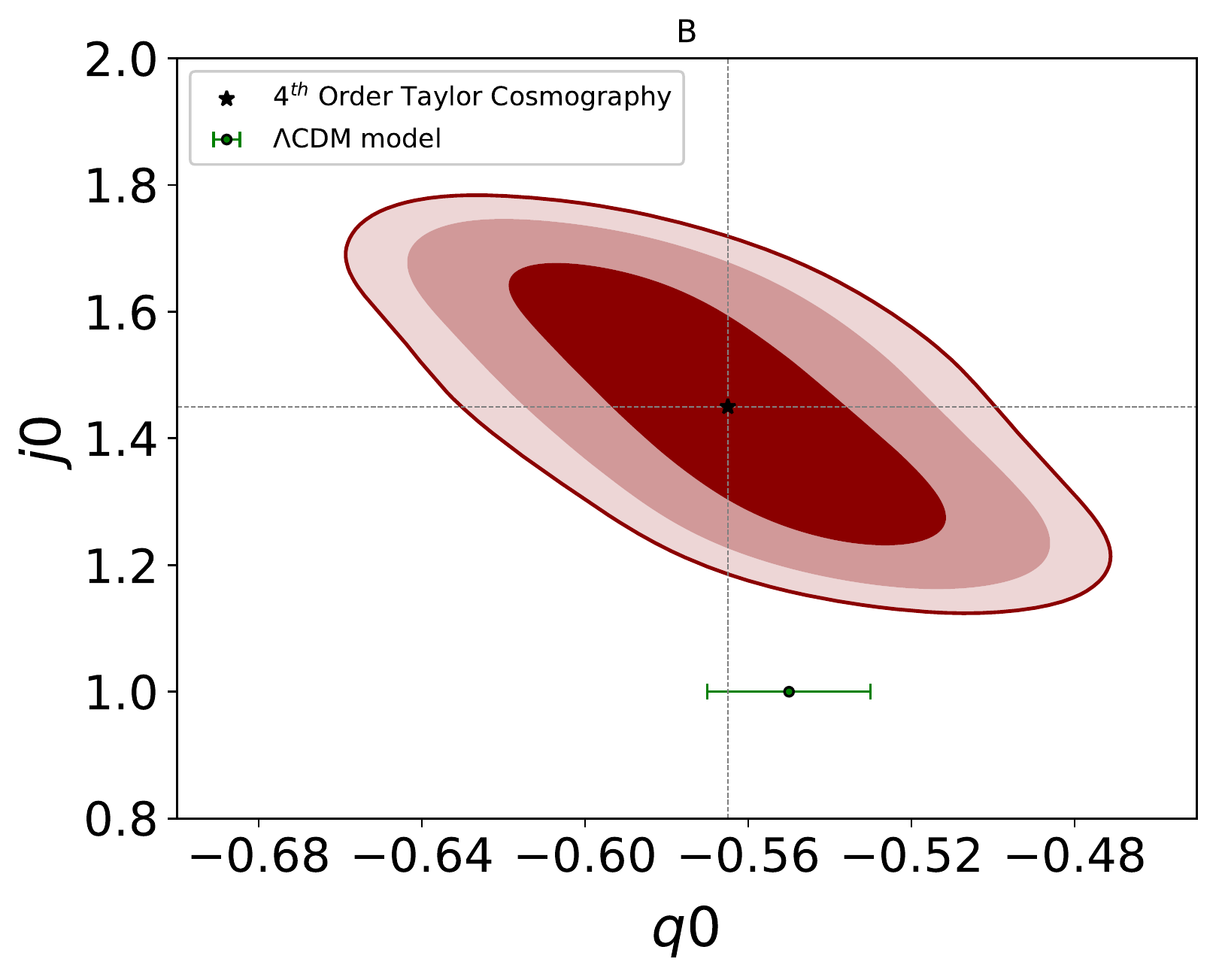}
	\includegraphics[width=7.6 cm]{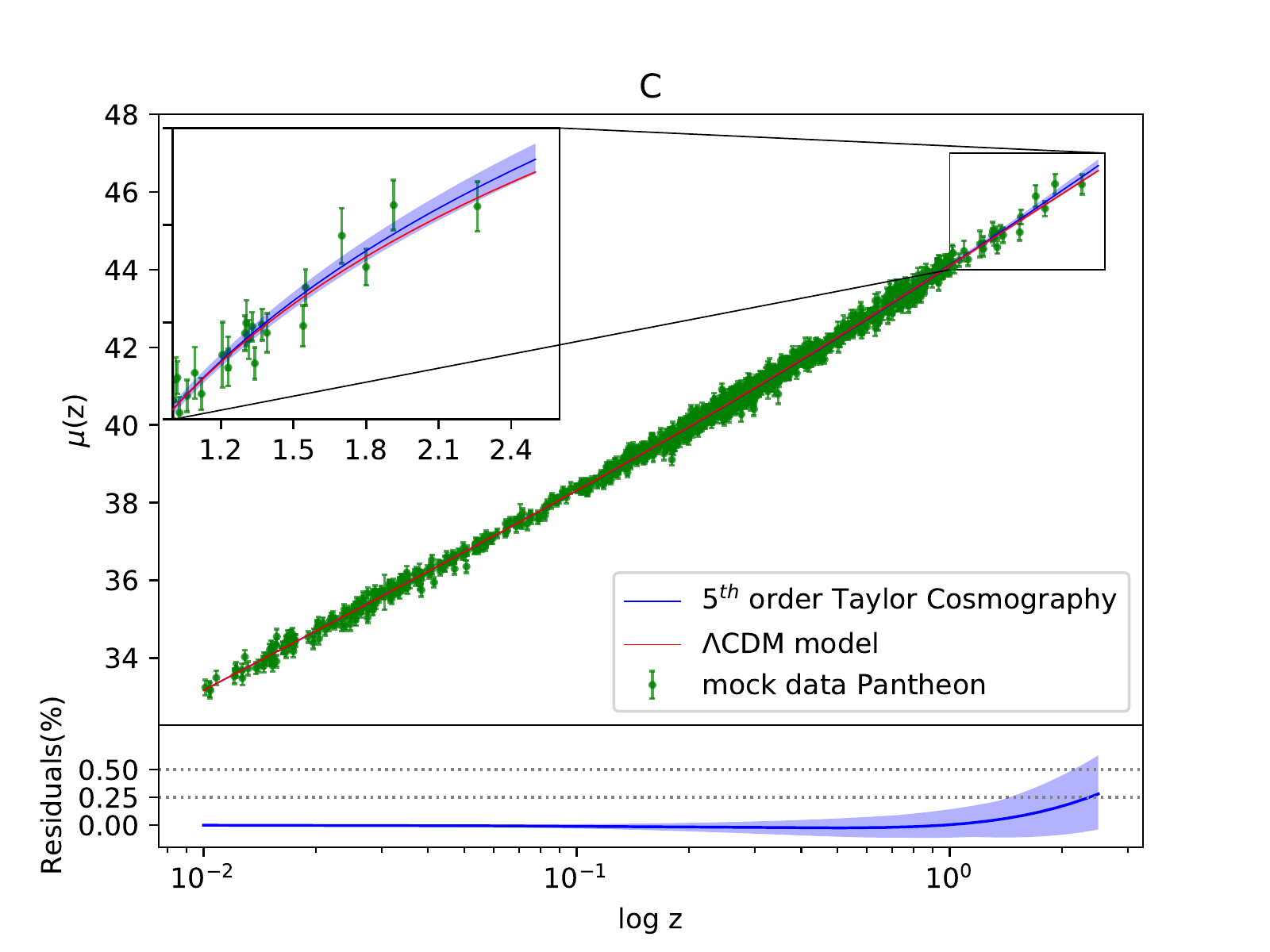}
	\includegraphics[width=6.8 cm]{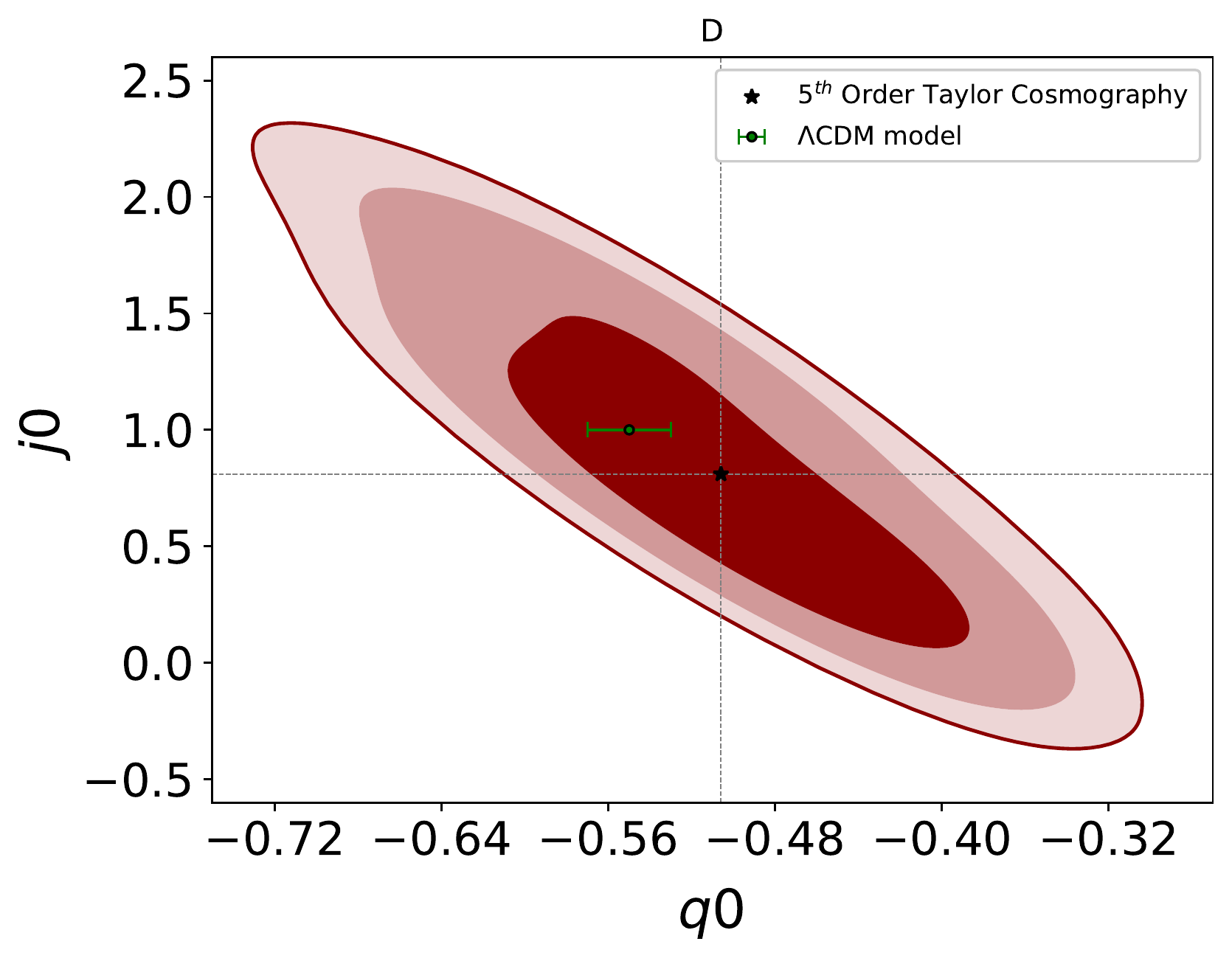}
	\includegraphics[width=7.6 cm]{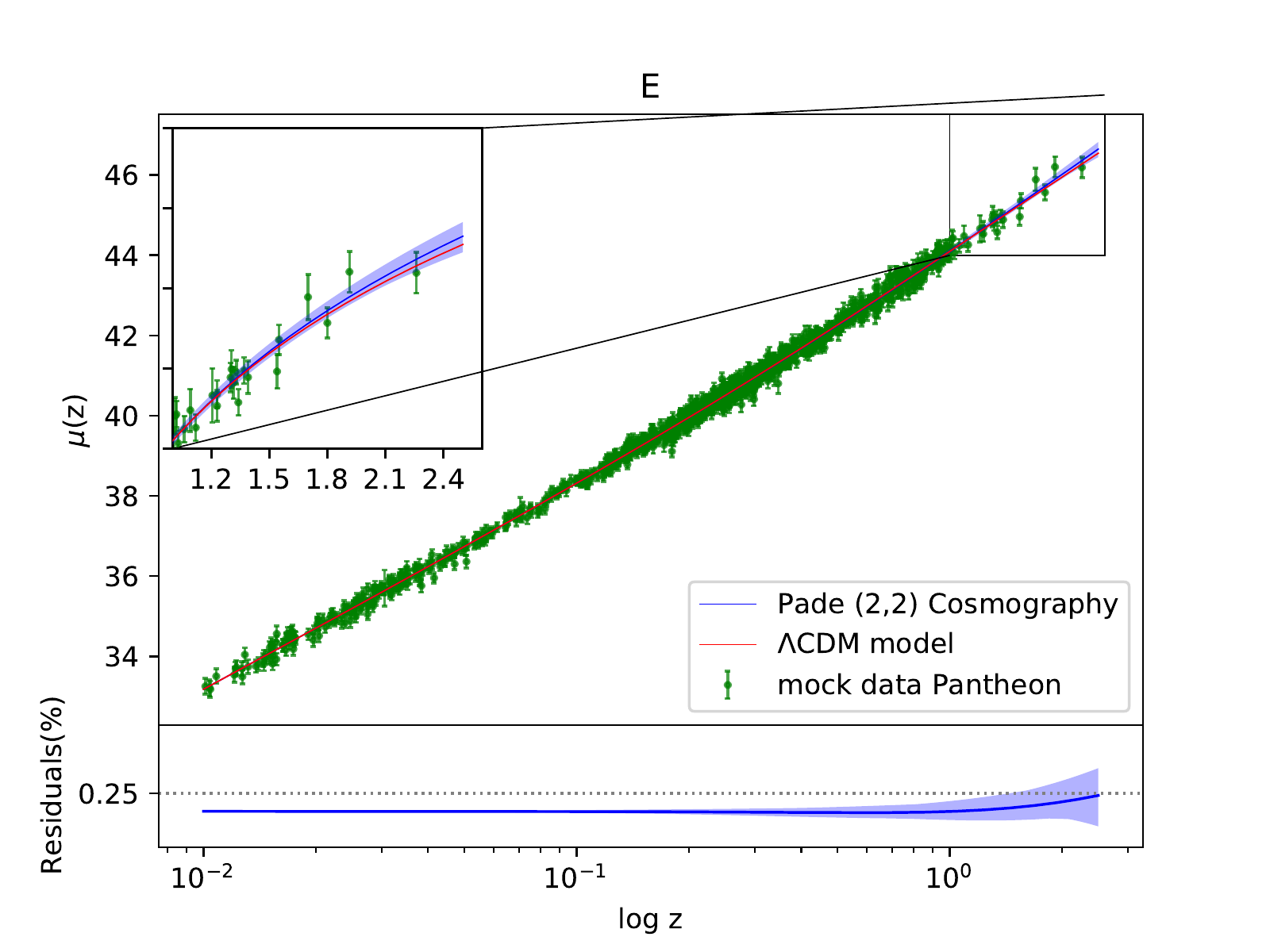}
	\includegraphics[width=6.8 cm]{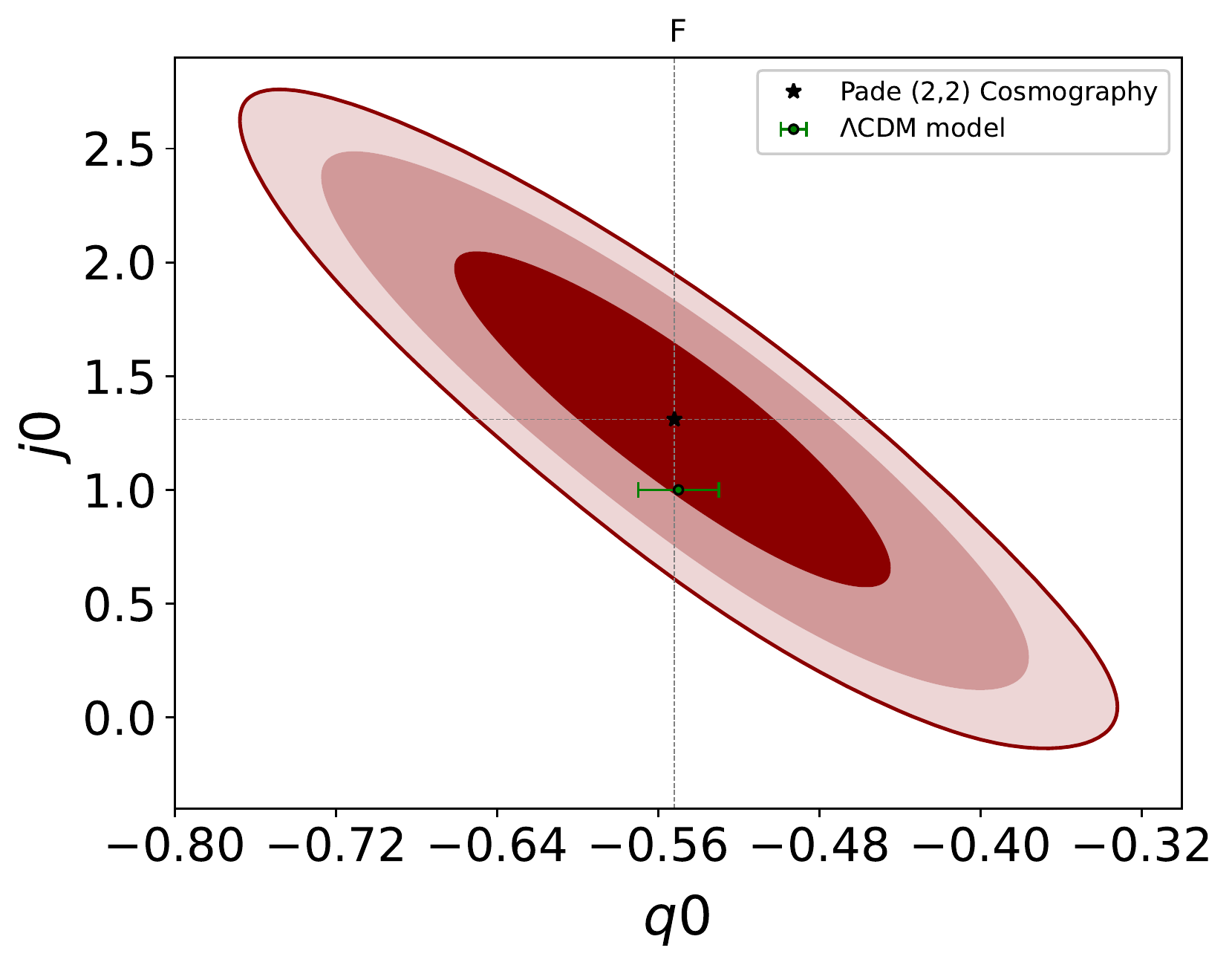}
	\includegraphics[width=7.6 cm]{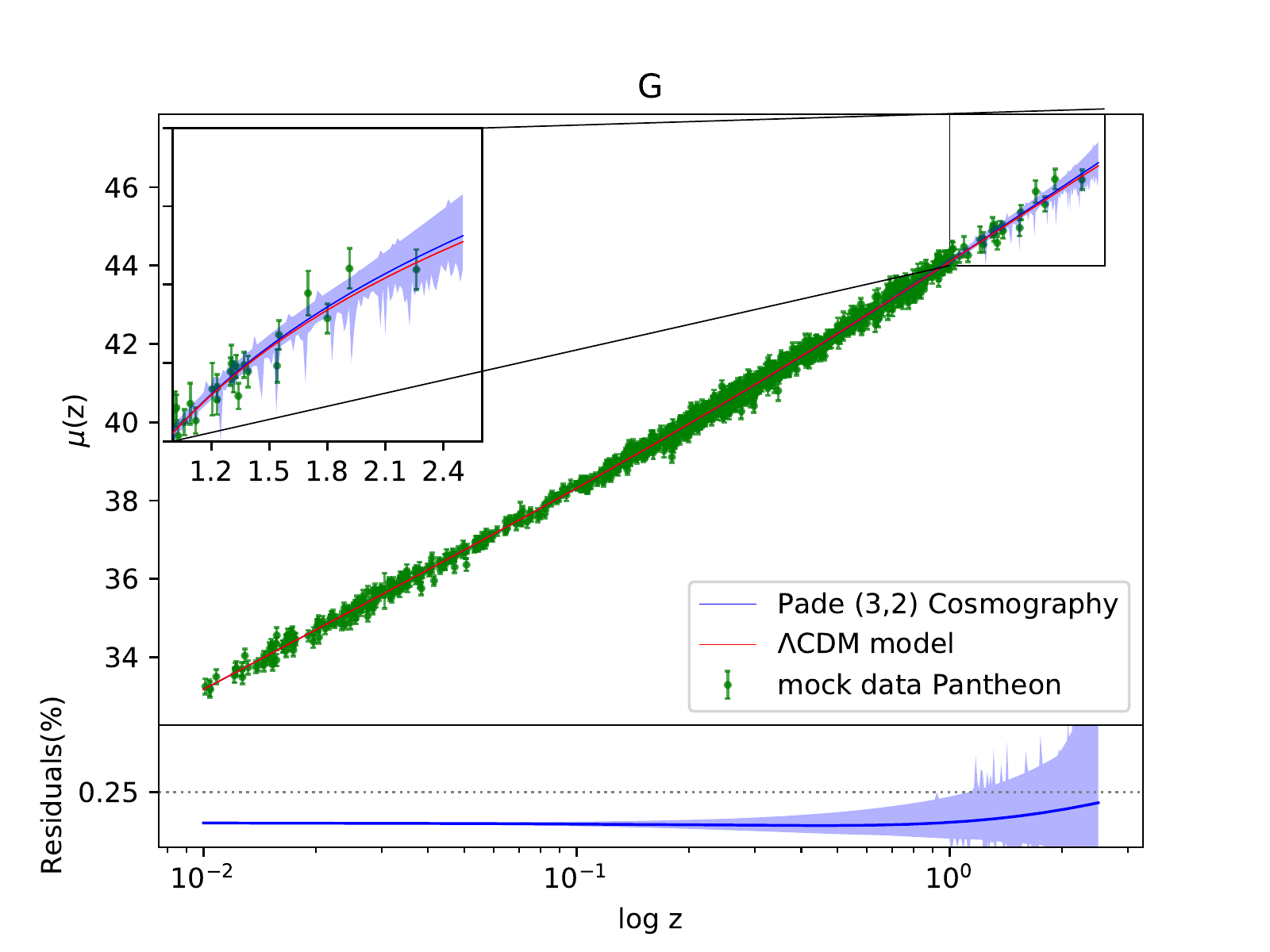}
	\includegraphics[width=6.8 cm]{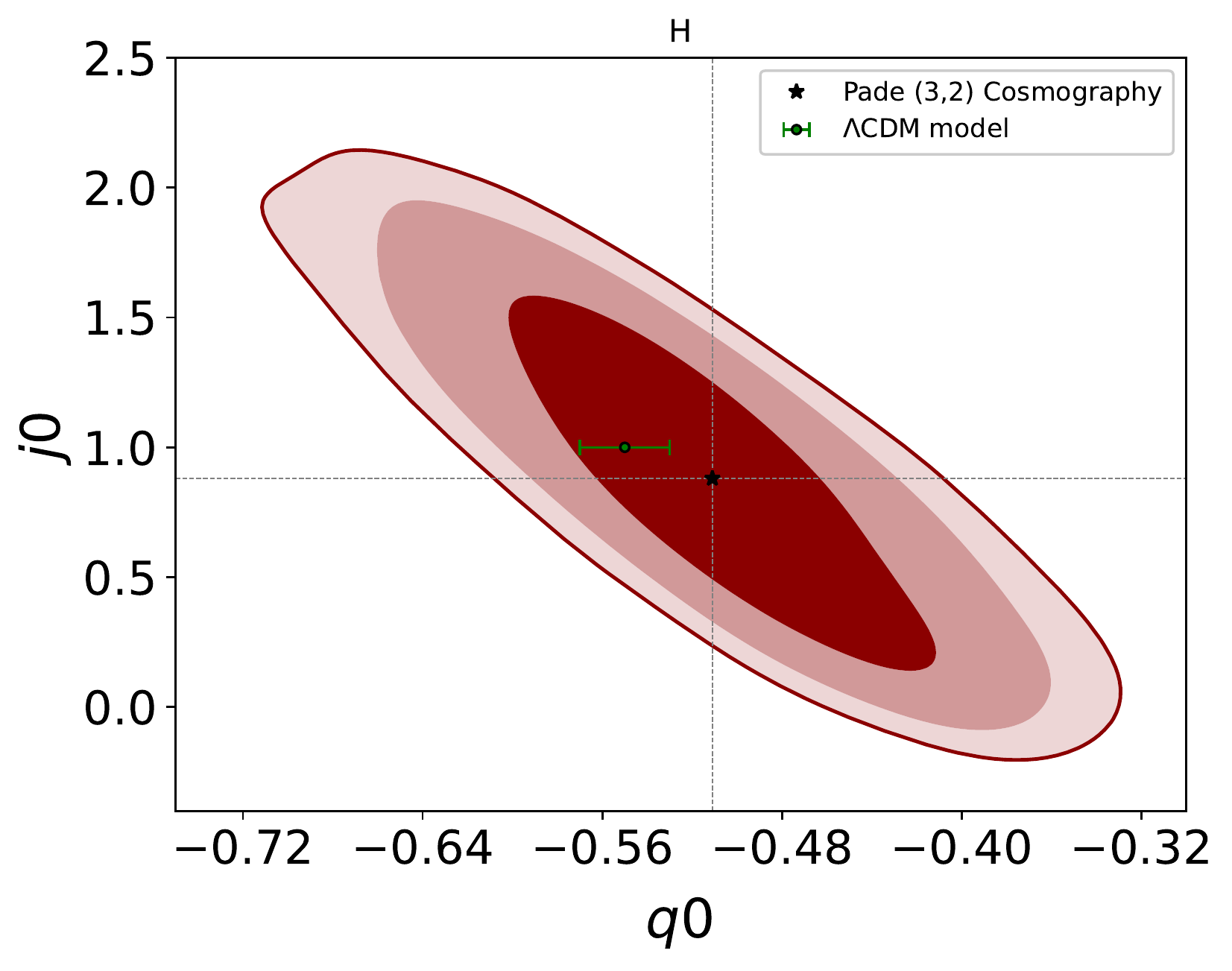}
	\caption{A comparison between  the $\Lambda$CDM model with the cosmographic method utilizing mock SNIa data. Left panels show the reconstructed distance modulus in the context of cosmographic method and its percentage difference with that of the standard $\Lambda$CDM model. Right panels show the difference between $\Lambda$CDM values of $q_0$ and $j_0$ with confidence regions of cosmographic approach.
	}
	\label{fig:Panth_Mock}
\end{figure*}

\subsection{Mock SNIa Sample }
In Fig.(\ref{fig:pantheon_LCDM}), our constrain utilizing mock SNIa data indicates $\Omega_{m0}= 0.3 \pm{0.013}$.
Since the canonical value of $\Omega_{m0}$ is in $1\sigma$ confidence level, we conclude that the mock data for the Hubble diagram of SNIa are properly generated on $\mu_{\Lambda}(z)$. Furthermore, using Eq.(\ref{eq17}), the best-fit values of the cosmographic parameters for the flat standard model are obtained as reported in the first row of Tab.\ref{tab:LCDM_Pan_Mock}. On the other hand, using mock SNIa data, we constraint the cosmographic parameters model-independently for different cosmographic methods defined based on the $4^{\it th}$ and $5^{\it th}$ order Taylor expansions as well as Pade (2,2) and Pade (3,2) approximations, respectively. Our numerical results for various cosmographic methods are presented in Tab. \ref{tab:Pan_Mocks}. In Fig.(\ref{fig:fig2}), we show $1\sigma$ to $3\sigma$ confidence levels of $q_{0}$ and $j_{0}$ parameters for the different cosmographic methods. We observe that in the cases of cosmographic methods beyond $4^{\it th}$ order Taylor series, we get larger confidence regions for cosmographic parameters $q_0$ and $j_0$. Now, using the best-fit values of cosmographic parameters, we reconstruct the distance modulus for different cosmographic methods and compare them with that of the standard flat-$\Lambda$CDM model. We also compare the best-fit values of the cosmographic parameters $q_0$ and $j_0$ obtained in flat-$\Lambda$CDM with the confidence regions of the same parameters obtained in cosmographic methods. Results are shown in Fig. \ref{fig:Panth_Mock}. Our results for the cosmographic method based on the $4^{\it th}$ order Taylor series are shown in panels (A \& B). In panel (A), we observe that the difference between the distance modulus of the standard model and that of the cosmographic method based on the $4^{th}$ order Taylor series reaches to $0.39\%$ at redshift $z=2.5$. In panel (B), we observe no significant difference between $\Lambda$CDM value and confidence region of cosmography for $q_0$ parameter, while a $3.2 \sigma$ deviation is seen for the $j_{0}$ parameter. We also observe tensions $6.22\sigma$ and $5.22\sigma$, respectively, for higher cosmographic parameters $s_0$ and $l_0$ by comparing first rows of Tabs. (\ref{tab:LCDM_Pan_Mock} \& \ref{tab:Pan_Mocks}). These big differences indicate that the cosmographic method based on the $4^{\it th}$ order Taylor series is inadequate to reconstruct the distance modulus up to redshift $z=2.5$. We emphasize that mock SNIa data were generated based on the standard $\Lambda$CDM model. Therefor, we expect the standard model to be well fitted to the mock data. On the other hand, as a model-independent approach, we expect the cosmographic method to be consistent with mock data and consequently with the standard $\Lambda$CDM model. Hence, in the analysis based mock data, any significant difference between the cosmographic approach and the standard model can be interpreted as an inadequacy of the cosmographic method due to errors of mathematical approximation. The insufficiency of the $4^{th}$ order Taylor expansion of the Hubble parameter is also confirmed from the statistical  AIC and BIC criteria in \cite{Hu:2022udt}.

 In panels (C \& D), we display the results for cosmography based on the $5^{\it th}$ order Taylor series. In panel (C), we observe that the percentage difference between the reconstructed distance modulus in cosmographic method and that of the flat-$\Lambda$CDM model reduces to $0.28\%$. Consequently, we detect no tension between flat-$\Lambda$CDM model and cosmographic method in $q_0-j_0$ plane in panel (D). Comparing the first row of Tab. \ref{tab:LCDM_Pan_Mock} and the second row of Tab.\ref{tab:Pan_Mocks}, we also observe that the higher cosmographic parameters $s_0$, $l_0$ and $m_0$ of the standard model are consistent with those of the cosmographic method. This means that we can apply the cosmographic method defined on the basis of the $5^{\it th}$ order Taylor series of the Hubble parameter for reconstructing the distance modulus at redshifts lower than $z=2.5$.
In panels (E \& F), we present our results for Pade (2,2) approximation. In this case the percentage difference reduces to the value $0.22\%$ and consequently we have no tension between cosmographic parameters $q_0$ and $j_0$ of the standard model and those of the cosmographic approach. In addition, we detect the statistical $1.3\sigma$ and $0.11\sigma$ errors, respectively, for the values of $s_{0}$ and $l_{0}$ parameters (compare first row of Tab.\ref{tab:LCDM_Pan_Mock} and third row of Tab.\ref{tab:Pan_Mocks}). Hence, cosmography based on the rational Pade (2,2) approximation is also an appropriate model-independent method for reconstructing the distance modulus diagram at SNIa redshifts. Notice that Pade (2,2) has one lower free parameter than $5^{\it th}$ order Taylor series. Though, Pade (2,2) approximation shows reasonable consistency with the $\Lambda$CDM model, examining Pade (3,2) approximation can also enhance the certainty of employing the Pade approximation in the context of the cosmographic approach. In panels (G \& H) of Fig. \ref{fig:Panth_Mock}, we present our results for Pade (3,2) cosmographic method. The difference between distance modulus of Pade(3,2) cosmography and the standard model is $0.16 \%$, which is the lowest value compared with the previous cases. It is reasonable to infer that in this case, the error of Pade (3,2) is so small that it can be considered almost negligible. On the other hand, same as Pade (2,2), the cosmographic parameters of the standard model, including $q_{0}$ and $j_{0}$ parameters, are well constrained in confidence levels of the cosmographic approach based the Pade (3,2) approximation. Correspondingly, the difference between the $s_{0}$, $l_{0}$, and $m_{0}$ parameters in the two approaches is $0.09 \sigma$, $1.08\sigma$ and $0.66 \sigma$, respectively. Thus, we have an excellent consistency between Pade (3,2) cosmographic method and standard model, indicating that this case of cosmographic method can be applied for reconstructing the low-redshift (redshifts smaller than $\sim2.5$) distance modulus diagram model-independently.\\

\begin{table*}
	\centering
	\caption{The best-fit value of $\Omega_{m0}$ for the standard $\Lambda$CDM model, using mock SNIa and QSOs data (left-part). The Best-fit values of cosmographic parameters of the standard $\Lambda$CDM model, using mock SNIa and QSOs data (Right Part).
	}
	\begin{tabular}{ c c c c c c c c c  }
		\hline \hline
		& $\Omega_{m0}$ & $\vert$ & $q_0$ &$j_0$ & $s_0$ & $l_0$ &$m_0$\\
		\hline
		Mock SNIa  &$0.300\pm 0.013$ & $\vert$ & $-0.550\pm 0.020$ &$1$ & $-0.351\pm 0.059$ & $3.12^{+0.14}_{-0.15}$ &$-10.92^{+0.92}_{-0.82}$\\
		\hline 
		Mock QSOs  & $0.2978\pm 0.0094$ &  $\vert$ & $-0.553\pm 0.014$ &$1$ & $-0.340\pm 0.042$ & $3.092^{+0.097}_{-0.11}$ &$-10.75^{+0.67}_{-0.58}$\\
		\hline \hline
	\end{tabular}\label{tab:LCDM_Pan_Mock}
\end{table*}

\begin{table*}
	\centering
	\caption{ Best-fit values and $1\sigma$ confidence intervals of the cosmographic parameters obtained from various cosmographic approaches, using mock SNIa data.
	}
	\begin{tabular}{c c c c c c c c c }
		\hline \hline
		&  & $q_0$ & $j_0$ & $s_0$ & $l_0$  & $m_0$ &  \\
		\hline
		& Taylor 4 & $-0.565\pm 0.033$ & $1.45\pm 0.14$ & $1.43^{+0.26}_{-0.30}$ & $1.52\pm 0.27$  & $-$ &  \\
		\hline
		& Taylor 5 & $-0.506^{+0.075}_{-0.068}$ & $0.81^{+0.40}_{-0.53}$ & $-0.9^{+1.2}_{-2.2}$ & $0.9^{+2.9}_{-3.4}$  & $-1.1^{+7.9}_{-6.7}$ & \\
		\hline
		& Pade (2,2) & $-0.552\pm 0.068$ & $1.31\pm 0.46$ & $1.3^{+1.5}_{-1.1}$ & $2.9^{+2.3}_{-1.5}$  & $-$ & \\
		\hline
		& Pade (3,2) & $-0.511\pm 0.064$ & $0.88\pm 0.45$ & $-0.2\pm 1.7$ & $0.3\pm 2.6$  & $-3^{+13}_{-11}$&  \\
		\hline \hline
	\end{tabular}\label{tab:Pan_Mocks}
\end{table*}

\begin{figure} 
	\centering
	\includegraphics[width=6.5 cm]{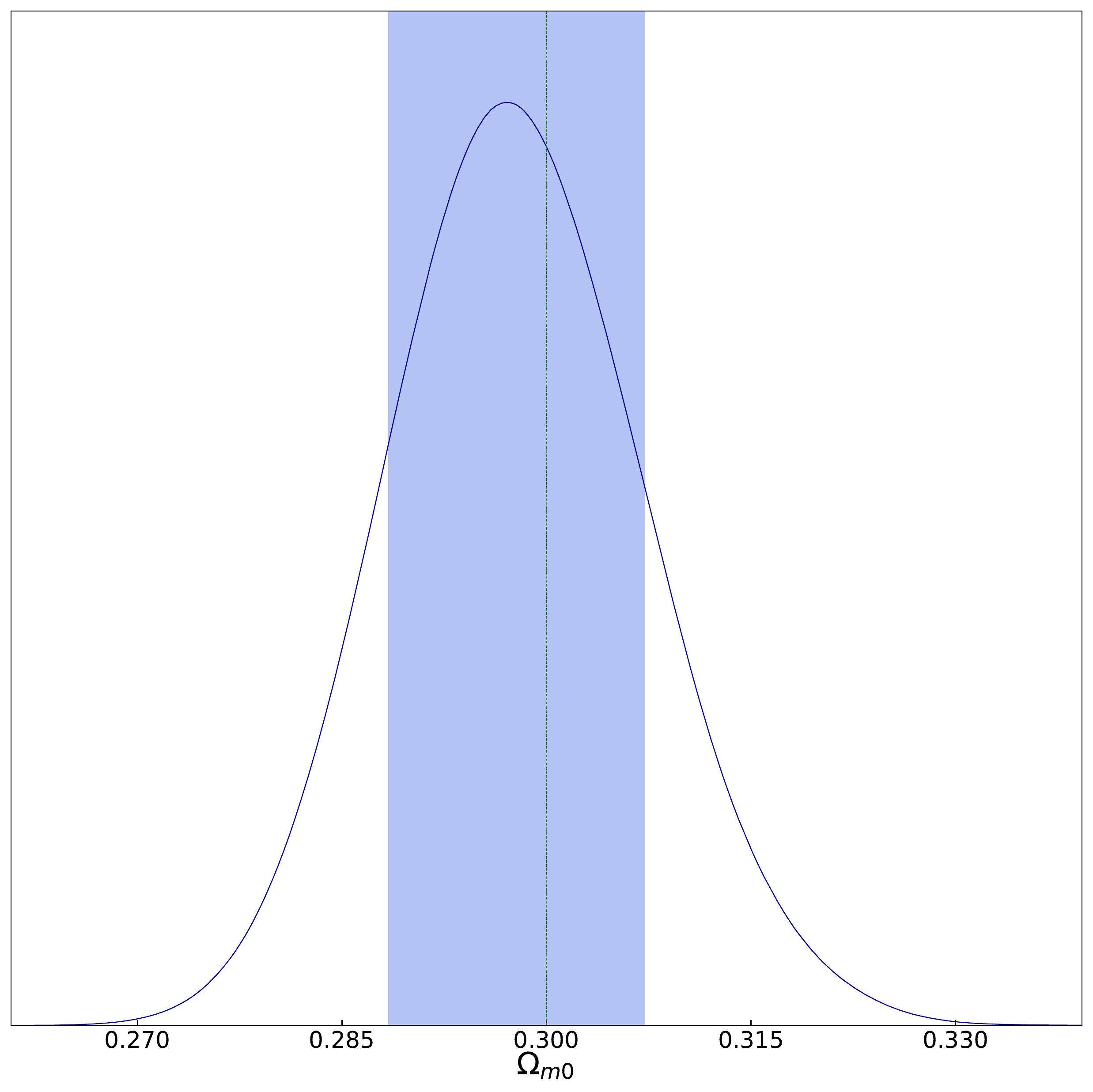}
	\caption{Same as Fig.(\ref{fig:pantheon_LCDM}), but for mock QSOs data.}
	\label{fig:quasar_LCDM}
\end{figure}
\begin{figure} 
	\centering
	\includegraphics[width=6.5 cm]{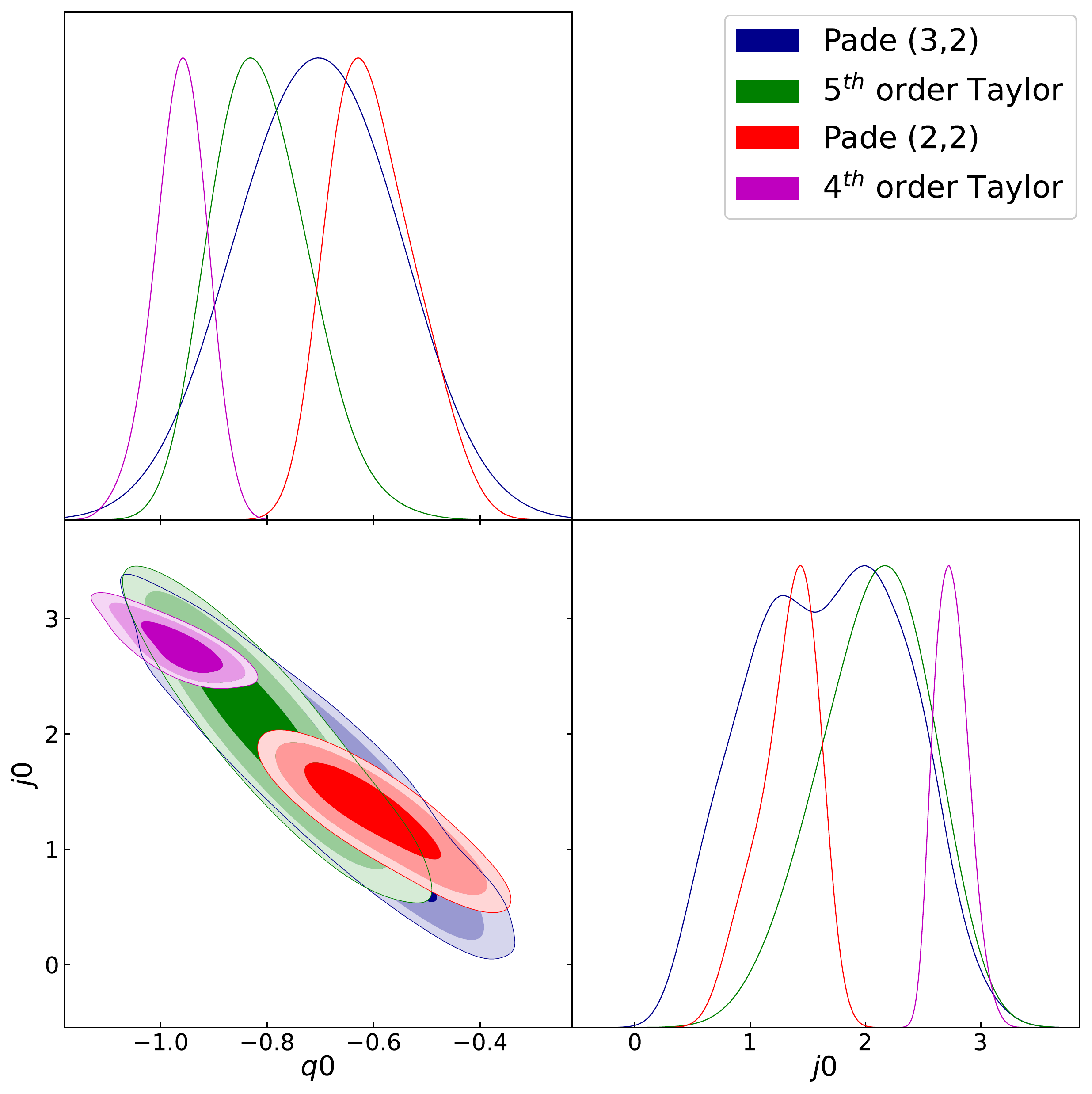}
	\caption{Same as figure \ref{fig:pantheon_LCDM} for quasars mock data.}
	\label{fig:fig5}
\end{figure}

\begin{figure*} 
	\centering
	\includegraphics[width=7.6 cm]{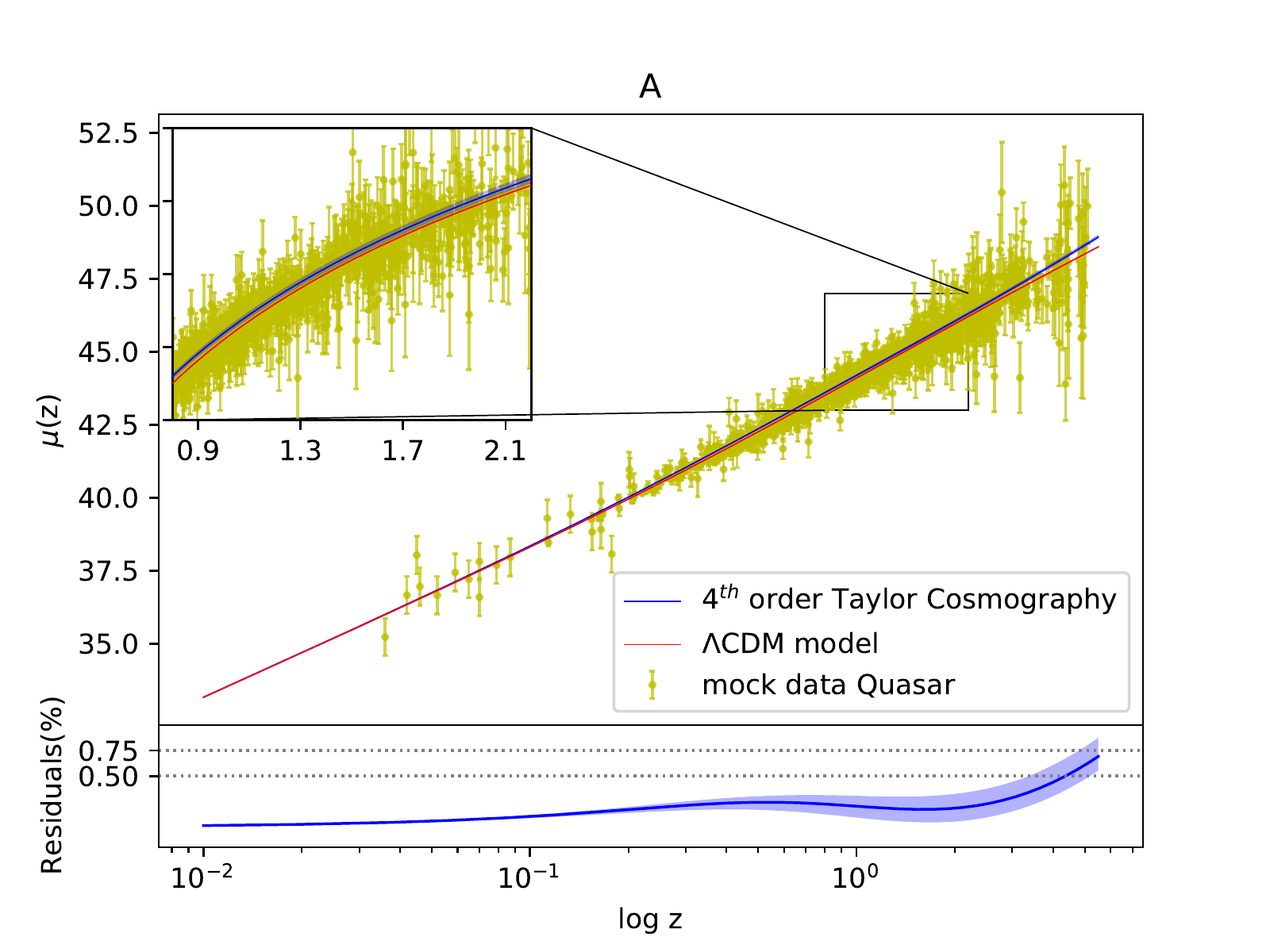}
	\includegraphics[width=6.8 cm]{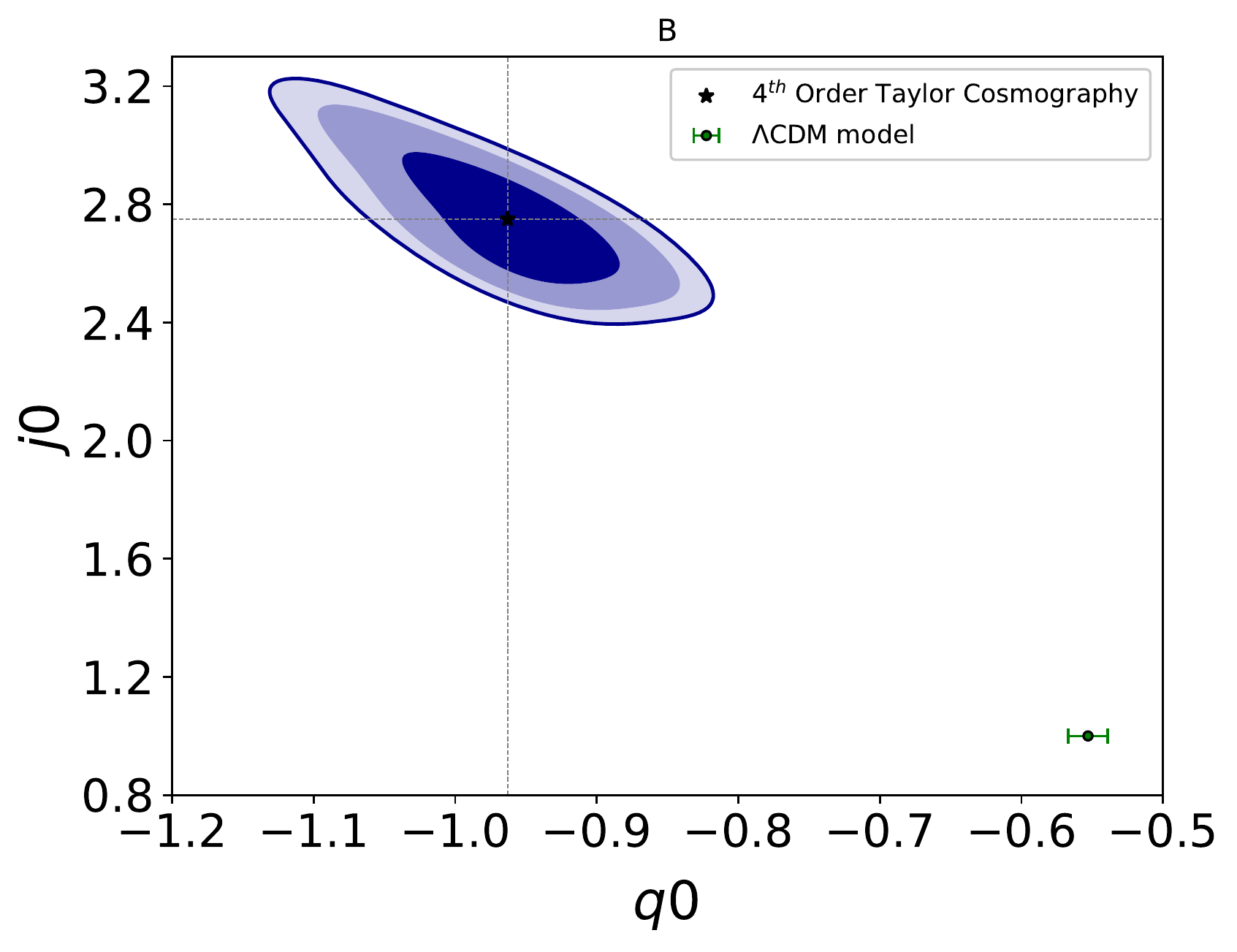}
	\includegraphics[width=7.6 cm]{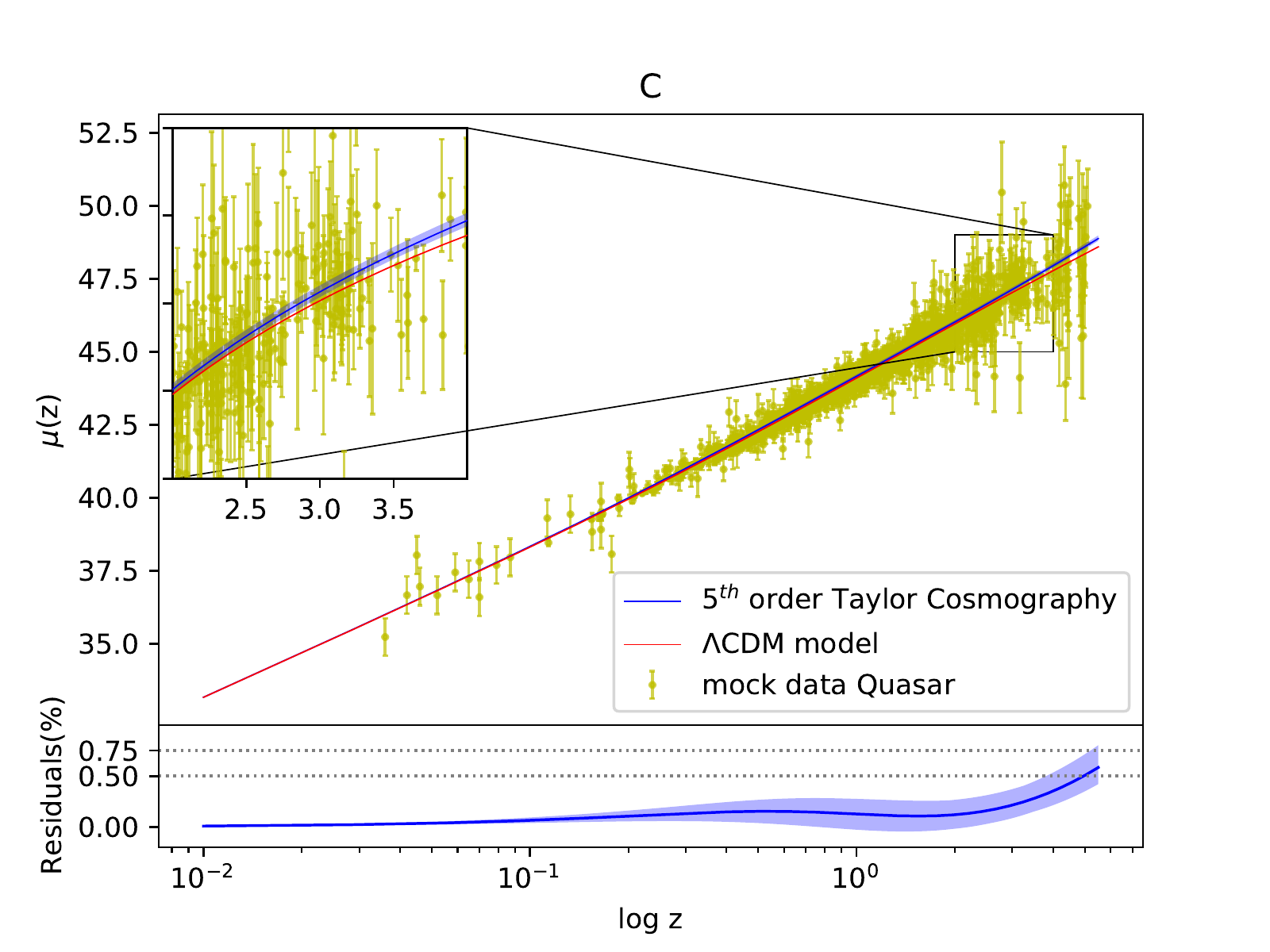}
	\includegraphics[width=6.8 cm]{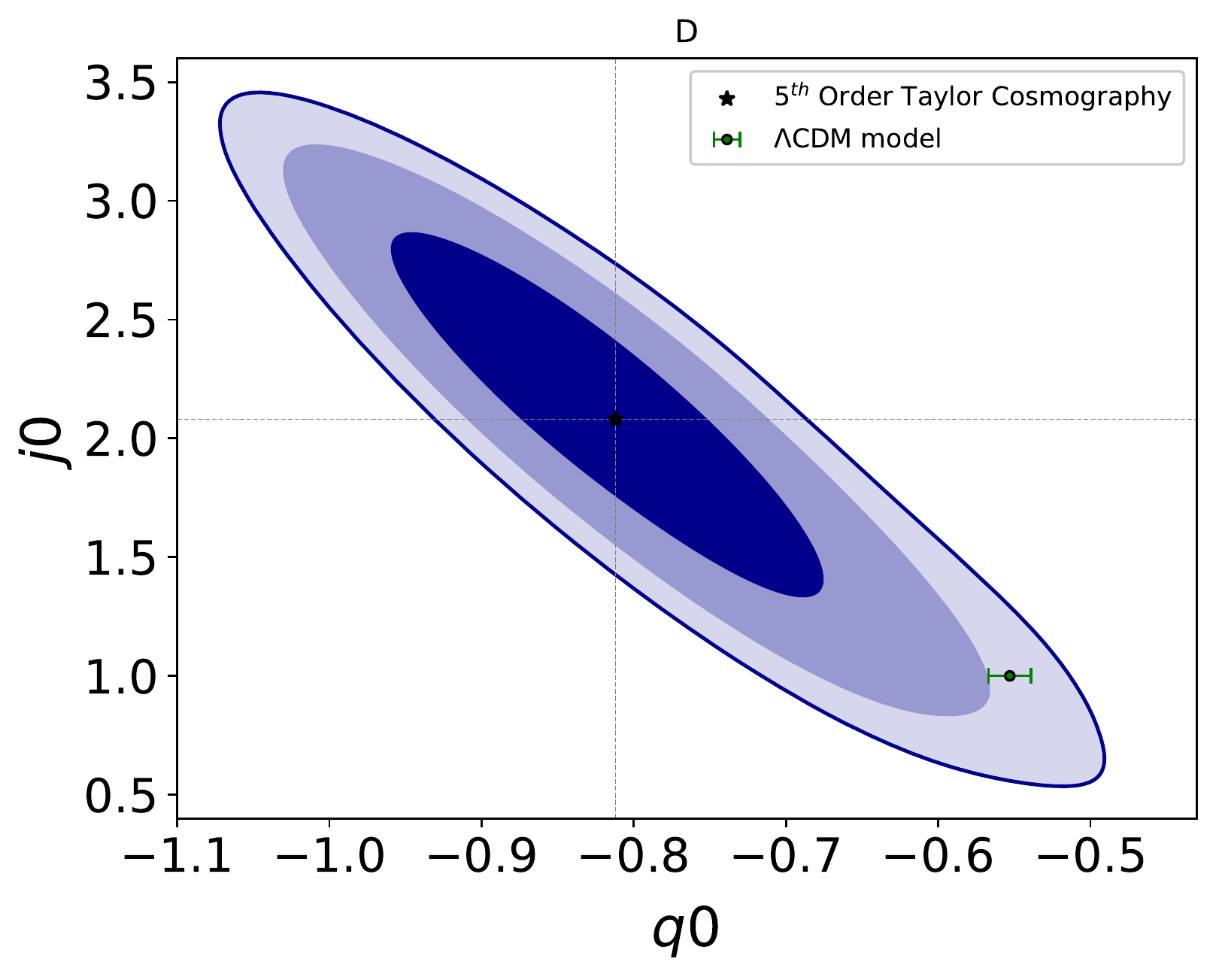}
	\includegraphics[width=7.6 cm]{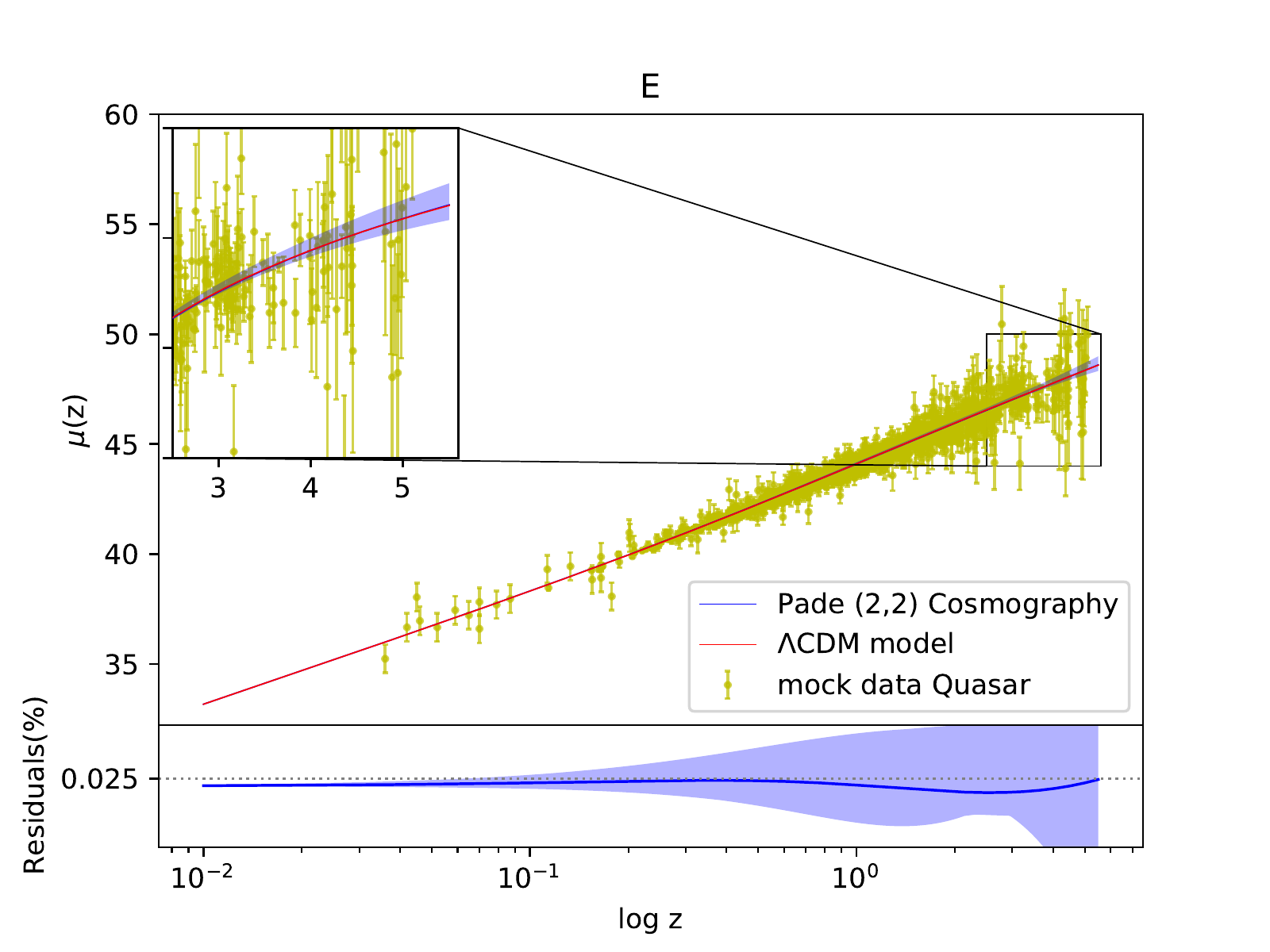}
	\includegraphics[width=7.2 cm]{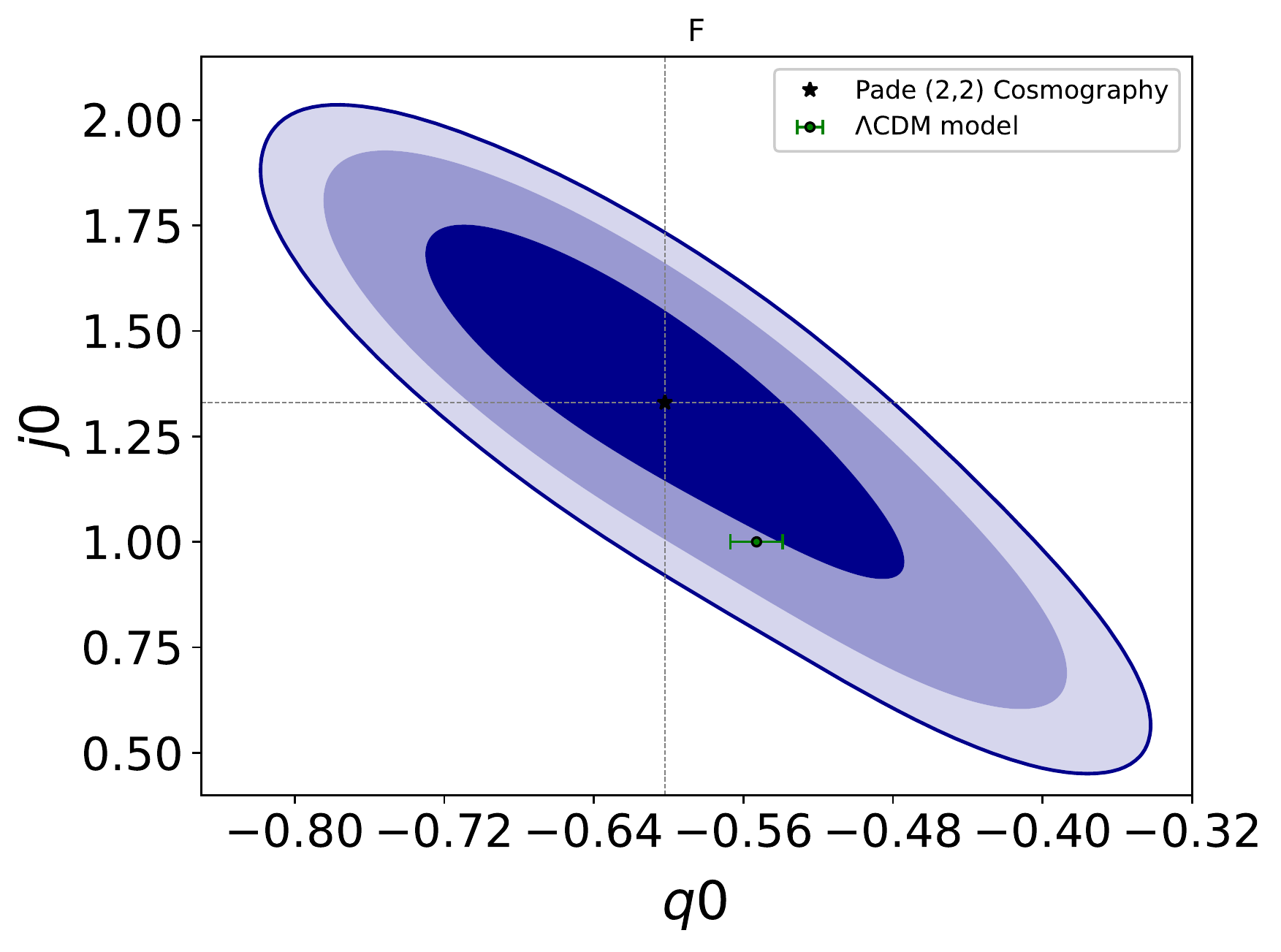}
	\includegraphics[width=7.6 cm]{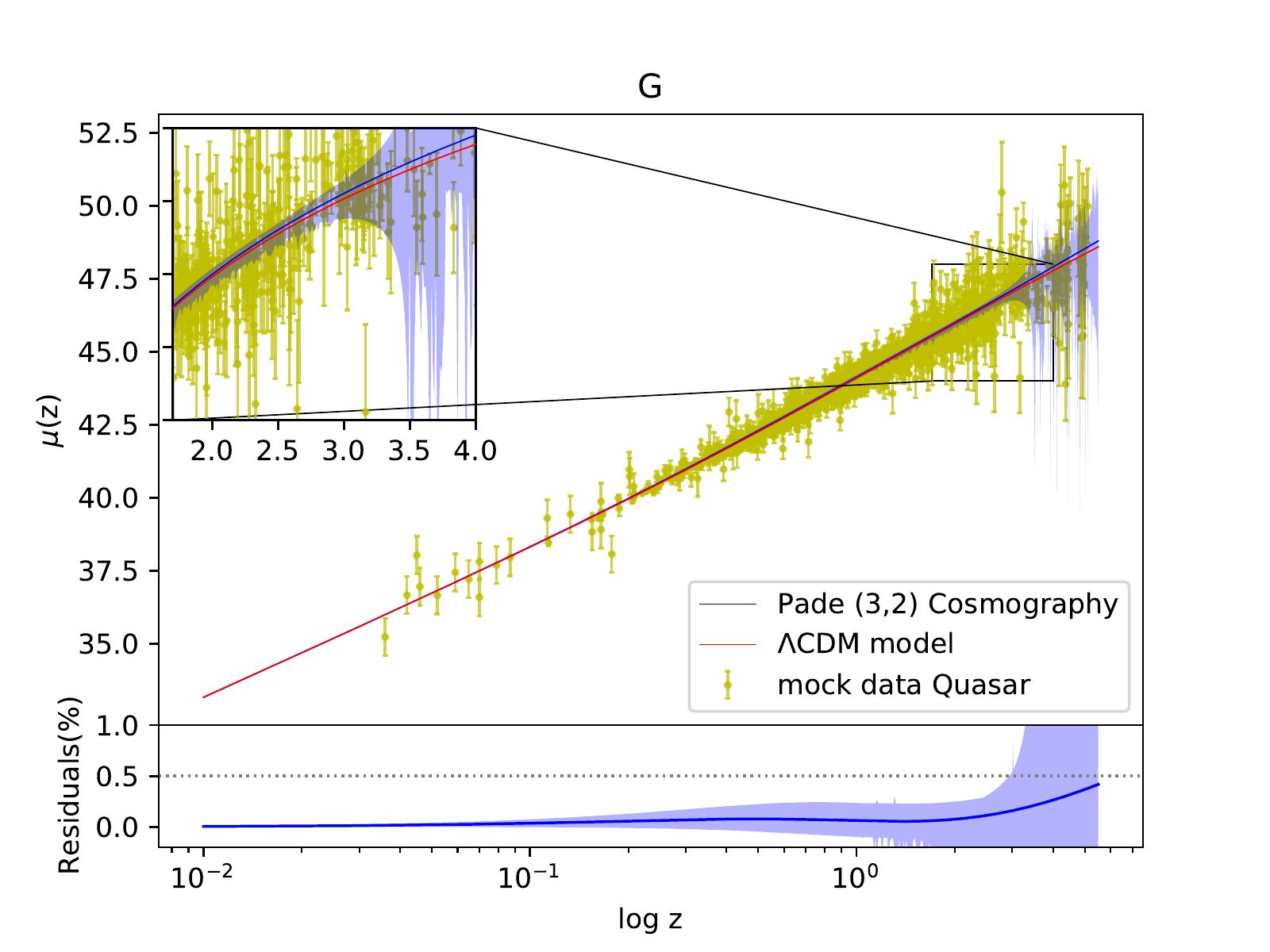}
	\includegraphics[width=7.2 cm]{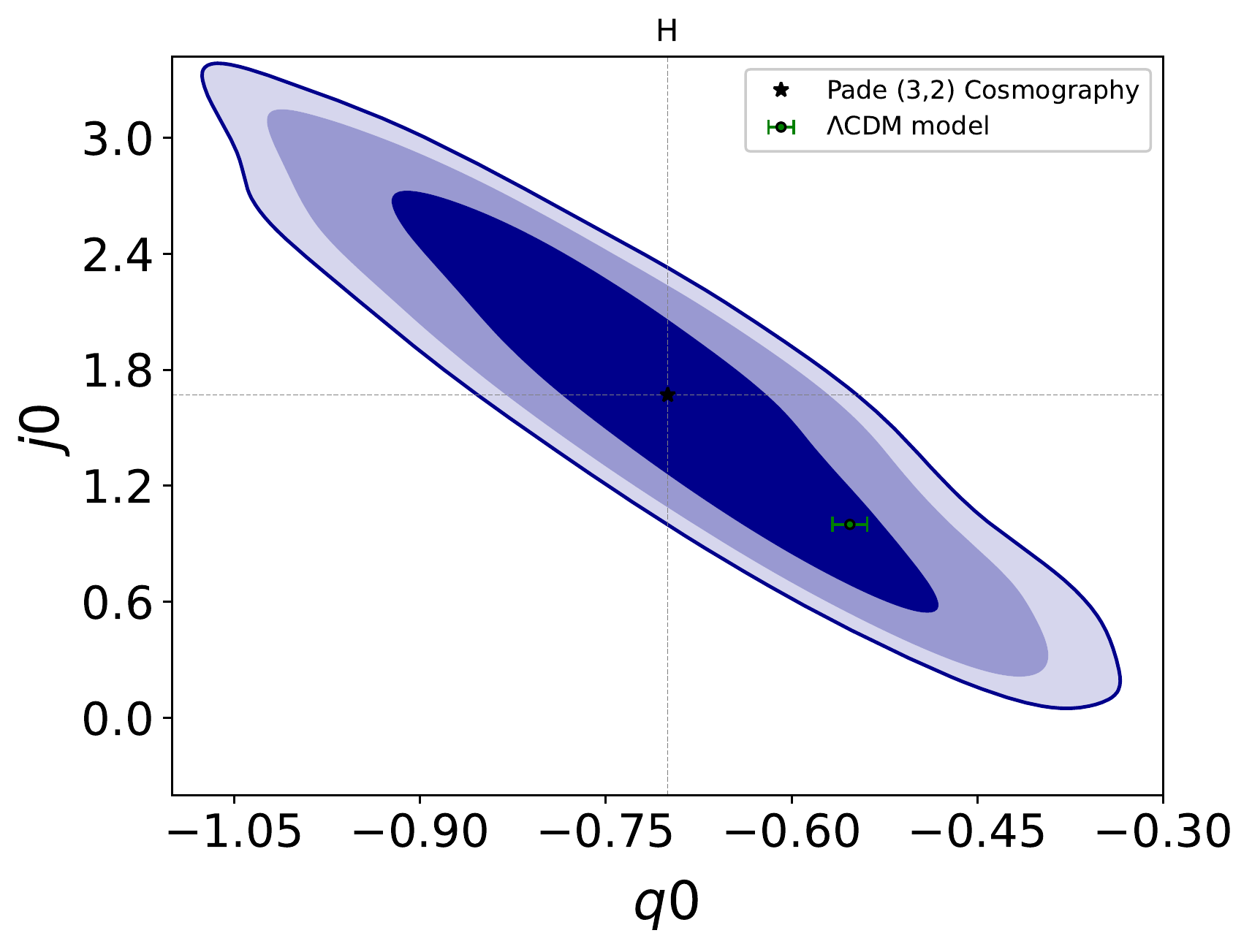}
	\caption{Same as Fig.(\ref{fig:Panth_Mock}), but for mock QSOs data.
	}
	\label{fig:quasar_Mock}
\end{figure*}
\subsection{Mock QSOs sample}

Here we present our numerical results using mock QSOs data for the Hubble diagram. Our procedure is the same as one presented in the previous sub-section. We obtain the constrained value of matter density for the flat-$\Lambda$CDM model as $\Omega_{m0}=0.2978 \pm{0.0094}$. In Fig. \ref{fig:quasar_LCDM}, we see that the canonical value $\Omega_{m0}=0.3$ is full consistent with the constrained value with the difference lower than $1\sigma$ error. This means that mock data for Hubble diagram of QSOs generated based on the canonical value $\Omega_{m0}=0.3$ and $h=0.7$ are properly fitted to the standard $\Lambda$CDM model. Now, using Eq. (\ref{eq17}) and constraints on $\Omega_{m0}$ obtained from mock QSO sample, we obtain the best fit-values and $1\sigma$ to $3\sigma$ confidence levels of cosmographic parameters in the flat-$\Lambda$CDM model, as reported in the second row of Tab.(\ref{tab:LCDM_Pan_Mock}). Comparing the first and second rows of  Tab.(\ref{tab:LCDM_Pan_Mock}), we see that the best-fit values of cosmographic parameters in flat-$\Lambda$CDM model respectively obtained from mock SNIa and mock QSOs are consistent to each other. In the next step, using mock QSOs data, we constrain the cosmographic parameters in the context of various cosmographic methods considered in this work. Results are presented in Tab.(\ref{tab:Quasar_Mocks}). In addition, in Fig.(\ref{fig:fig5}), we show $1\sigma$ to $3\sigma$ confidence regions of cosmographic parameters $q_0$ and $j_0$ for different cosmographic methods. As a quick result, we observe that the confidence level of $j_0$ obtained from the cosmographic method based on the $4^{\it th}$ order of Taylor expansion completely differs from $j_0=1$ which means that this cosmographic method is not valid. Using the best-fit values of the cosmographic parameters presented in Tab.(\ref{tab:Quasar_Mocks}), we reconstruct the Hubble diagram in the context of cosmographic methods as shown in left panels of Fig. (\ref{fig:quasar_Mock}). In right panels, we show the confidence regions of $q_0$ and $j_0$ parameters and compare the results of cosmographic methods with $\Lambda$CDM values. In panel (A), we observe that the percentage difference between reconstructed distance modulus in cosmographic method based on $4^{\it th}$ order Taylor series and $\mu_{\Lambda}$ reaches to $0.69\%$ at $z\sim 5.5$. In panel (B), our constraint shows that $q_{0}$ and $j_{0}$ parameters in cosmographic approach based on the $4^{\it th}$ order Taylor series are equal to $q_0=-0.963^{+0.054}_{-0.046}$ and $j_0=2.75^{+0.13}_{-0.17}$, while corresponding values in the standard flat-$\Lambda$CDM model are $q_0=-0.553^{+0.014}_{-0.014}$ and $j_0=1.00$. This issue indicates a big difference equal to $7.9\sigma$ on $q_{0}$ and $11.6\sigma$ on $j_{0}$ parameters, as shown in panel (B). According to the above result, it is evident that the cosmographic approach based on the $4^{\it th}$ order Taylor expansion of the Hubble parameter will not be consistent with the standard model at higher redshift \citep[see also][]{Risaliti:2018reu, Rezaei:2020lfy}. This inconsistency is due to the error of $4^{\it th}$ order Taylor series at high-redshifts.  Nevertheless, as shown in panel (C) of Fig.(\ref{fig:quasar_Mock}), utilizing the cosmographic approach based on the $5^{\it th}$order Taylor expansion of the Hubble parameter can decrease the percentage difference to the value $0.58\%$. In addition, we can notice that despite the Taylor's $4^{\it th}$ order expansion, no significant difference can be shown between $q_{0}$ and $j_{0}$ parameters in $\Lambda$CDM model and confidence regions in cosmographic method based on the $5^{\it th}$ order Taylor expansion, as shown in panel (D) of Fig.(\ref{fig:quasar_Mock}). It can be concluded that by increasing the order of Taylor expansion, the difference between cosmographic parameters of the $\Lambda$CDM model and confidence regions of the cosmographic method based on the Taylor series will decrease. In the other words, the inconsistency between Taylor cosmographic method and the standard model will vanish. Comparing the constrained values of the second row of Tab.(\ref{tab:Quasar_Mocks}) and the second row of Tab.(\ref{tab:LCDM_Pan_Mock}), we observe no significant difference between higher cosmographic parameters of the model-independent cosmographic approach and flat-$\Lambda$CDM cosmology. This result shows that we can use the cosmographic method based on the $5^{\it th}$ order Taylor expansion to reconstruct the distance modulus up to high-redshift $z\sim 5$. In panels (E \& F) of Fig.(\ref{fig:quasar_Mock}), our results for the cosmographic method based on the Pade (2,2) approximation have been shown. As shown in panel (E), employing cosmography based on the Pade (2,2) approximation reduces the difference in distance modulus diagrams to $0.02\%$. In addition, panel (F) shows that in this case, the consistency between the cosmographic parameters in the model-independent approach and the standard model is higher than the previous case. Ultimately, the cosmographic parameters have been constrained using the cosmographic approach based on the Pade (3,2) approximation. In this case, the difference between the distance modulus diagram in the standard model and the cosmographic approach is $0.42\%$, which is less than the $4^{\it th}$ and $5^{\it th}$ order Taylor expansions but more than Pade (2,2) approximation. According to the panels G and H in Fig.(\ref{fig:quasar_Mock}), together with the results of fourth-row in Tab.(\ref{tab:Quasar_Mocks}), we report no deviation between the cosmographic parameters in this approach and the standard model. Based on the above results, it can be inferred that in higher redshifts, cosmography using the $5^{\it th}$ order Taylor expansion and the approximations of the Pade(2,2) and (3,2) can be a reliable approach for reconstructing the Hubble diagram of cosmological objects like QSOs and GRBs. Although the $5^{\it th}$ order Taylor series in cosmographic method works correctly at high-redshifts, the Pade (2,2) approximation with one lower free parameter is more reliable approximation to be employed. Accordingly in the next section, we constrain the cosmographic parameters using the real observational data for the Hubble diagrams of SNIa, QSOs and GRBs in the context of cosmographic method defined based on the Pade (2,2) and Pade (3,2) approximations. In this concern, if the observational data reveals any significant deviation between the cosmographic parameters of cosmographic methods and flat-$\Lambda$CDM model, it can be interpreted as an observational tension for standard model.

\begin{table*}
	\centering
	\caption{Same as Tab.(\ref{tab:Pan_Mocks}), but using mock QSOs data.
	}
	\begin{tabular}{c c c c c c c c c }
		\hline \hline
		&  & $q_0$ & $j_0$ & $s_0$ & $l_0$  & $m_0$ &  \\
		\hline 
		& Taylor 4 & $-0.963^{+0.054}_{-0.046}$ & $2.75^{+0.13}_{-0.17}$ & $0.15^{+0.78}_{-0.24}$ & $0.01\pm 0.57$  & $-$ &  \\
		\hline
		& Taylor 5 & $-0.812^{+0.082}_{-0.10}$ & $2.08^{+0.54}_{-0.45}$ & $ 0.19^{+2.0}_{-0.88}$ & $3.3^{+1.9}_{-1.5}$  & $-5.2^{+4.8}_{-8.8}$ &  \\
		\hline
		& Pade (2,2) & $-0.602^{+0.069}_{-0.095}$ & $1.33^{+0.32}_{-0.20}$ & $0.49^{+0.58}_{-0.44}$ & $4.80^{+1.1}_{-0.58}$  & $-$ &   \\
		\hline
		& Pade (3,2) & $-0.70\pm 0.14$ & $1.67\pm 0.66$ & $0.0\pm 1.6$ & $2.69^{+3.2}_{-0.94}$  & $-1.8\pm 7.1$ &  \\
		\hline \hline
	\end{tabular}\label{tab:Quasar_Mocks}
\end{table*}

\begin{table}
	\centering
	\caption{Upper part:Best-fit values of the cosmological parameters for the $\Lambda$CDM model, utilizing two data combinations consist of sample ({\it i}) and sample ({\it ii}). Middle part: Best-fit values of the cosmographic parameters of the standard $\Lambda$CDM model, using samples ({\it i}) \& ({\it ii}). Lower part: The minimum value of $\chi^2$ function and corresponding AIC value obtained for the standard $\Lambda$CDM cosmology.}
	\begin{tabular}{ c c c c c c c c c c }
		\hline \hline
	Cosmological parameters	& sample ({\it i}) & sample ({\it ii}) \\
		\hline
		$\Omega_{b0}$ & $0.0507\pm 0.0041$ &$-$  \\
		\hline
		$\Omega_{dm0}$& $0.2504\pm 0.0093$ &$-$  \\
		\hline
		$\Omega_{m0}$& $0.3011\pm 0.0080$ &$0.295\pm 0.012$  \\
		\hline
		$h$& $0.7167\pm 0.0018$ &$-$ \\
		%\hline
		%$\chi^{2}_{Best}$& $1183.53$ &$1221.11$ \\
		\hline
		Cosmographic parameters\\
		 \hline
		$q_0$ & $-0.548\pm 0.012$ &$-0.557\pm 0.019$  \\
		\hline
		$j_0$ & $1$ &$1$  \\
		\hline
		$s_0$& $-0.355\pm 0.036$ &$-0.328\pm 0.056$  \\
		\hline
		$l_0$& $3.128\pm 0.089$ &$3.06^{+0.13}_{-0.14}$  \\
		\hline
		$m_0$   & $-10.97\pm 0.54$ &$-10.57^{+0.86}_{-0.76}$ \\
		\hline 
		Statistical criteria\\
		\hline
		$\chi^2_{min}$   & $1148.49$ & $1220.12$ \\
		\hline
		AIC   & $1154.49$ & $1222.12$ \\
		\hline\hline
	\end{tabular}\label{tab:LCDM_real}
\end{table}

\begin{table}
	\centering
	\caption{Up: Best-fit values of the cosmographic parameters of the Pade cosmographic method obtained from minimizing $\chi^2$ function in the context of MCMC algorithm, using the observational data in sample ({\it i}). Down: The minimum value of $\chi^2$ function, the corresponding AIC values, and the difference between AIC value of Pade cosmographic approaches and that of the standard $\Lambda$CDM Universe obtained using sample ({\it i})}.
	\begin{tabular}{ c c c c c c c c c c }
		\hline \hline
		Cosmographic parameters	& Pade (2,2) & Pade (3,2) \\
		\hline
		$q_0$ & $-0.628^{+0.066}_{-0.045}$ &$-0.685^{+0.069}_{-0.062}$  \\
		\hline
		$j_0$ & $1.36^{+0.29}_{-0.45}$ &$1.59\pm 0.38$  \\
		\hline
		$s_0$& $0.13^{+0.39}_{-0.43}$ &$-1.14^{+0.59}_{-0.69}$  \\
		\hline
		$l_0$& $4.55^{+0.37}_{-0.25}$ &$0.0\pm 1.4$  \\
		\hline
		$m_0$   & $-$ &$-2.7^{+4.8}_{-7.8}$ \\
	%	\hline
	%	$\chi^{2}_{Best}$   & $1144.17$ &$1144.78$ \\
		\hline 
		Statistical criteria\\
		\hline
		$\chi^2_{min}$ & $1144.17$ &	$1142.75$  \\
		\hline
		AIC & $1152.17$ &	$1152.75$  \\
		\hline
		$\Delta AIC =AIC-AIC_{\Lambda CDM}$ & $-2.32$ & $-1.74$  \\
		\hline\hline
	\end{tabular}\label{tab:pade_lowz}
\end{table}

\begin{figure} 
	\centering
	\includegraphics[width=6.5 cm]{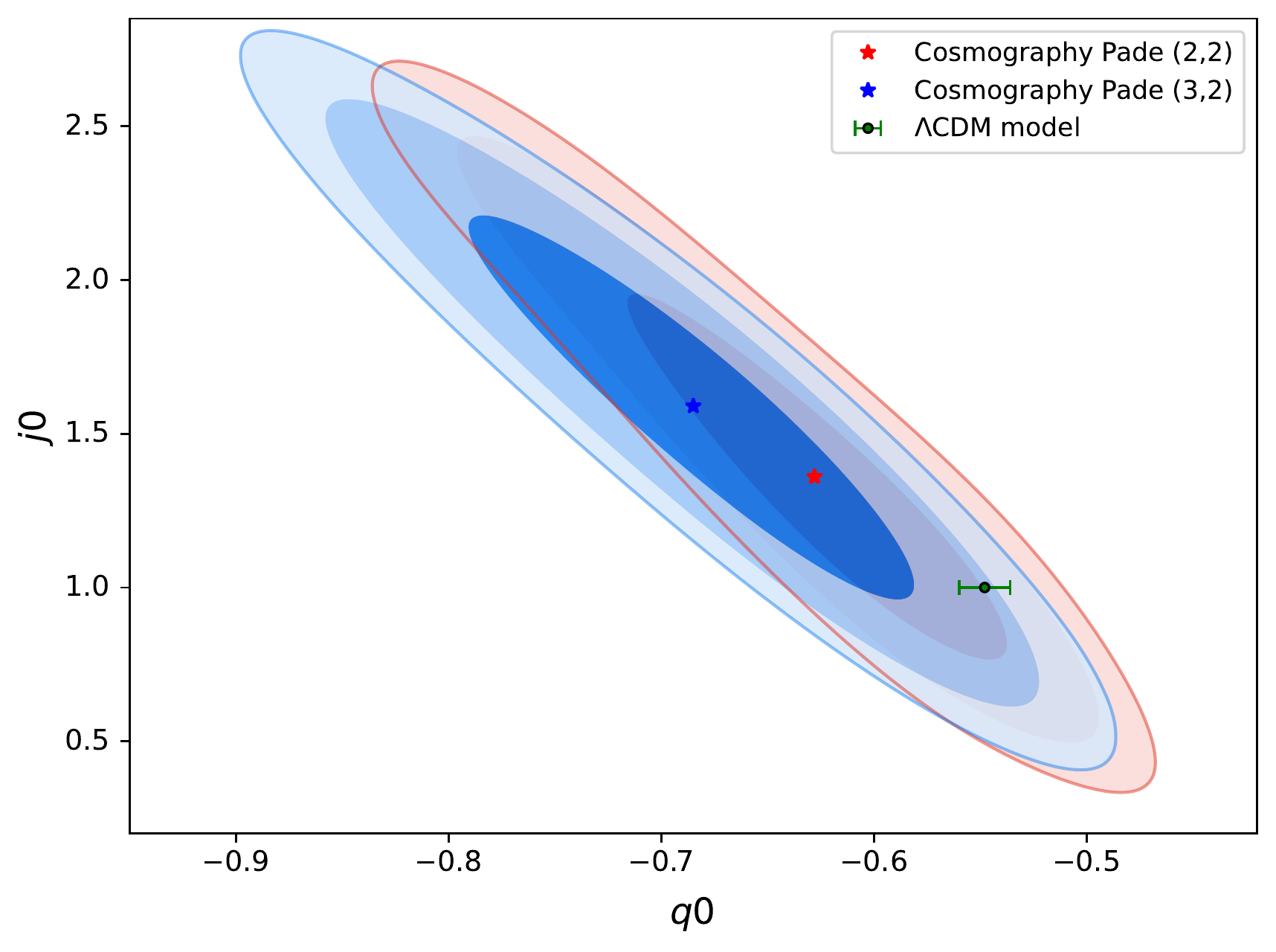}
	\caption{$1\sigma$ to $3\sigma$ confidence regions of the cosmographic parameters $q_0$ and $j_0$ obtained in the context of Pade (2,2) and Pade (3,2) cosmographic approaches using the combination of observational datasets in sample ({\it i}). The $\Lambda$CDM constraint for $q_0$ using the same datasets is shown for comparison.
	}
	\label{fig:real_low}
\end{figure}

\section{Observational data} \label{sect:Real_Data}
As shown in the previous section, the Pade (2,2) and (3,2) polynomials, despite Taylor's $4^{\it th}$ order expansion, can reconstruct the proper distance modulus in the model-independent approach at both redshifts $z<2.5$ and $z<5.5$. In this section, we compare the flat-$\Lambda$CDM model with the Pade cosmographic approaches, using the real observational data. Since the error truncation of Pade (2,2) and Pade (3,2) are negligible at least up to redshift $z\sim 5.5$, if any tension between the cosmographic parameters of standard model and those of the cosmographic method is revealed using the observational Hubble diagrams or other cosmological data, it is due to the standard model itself. In this study, we set two combinations of observational data samples including Hubble diagram of SNIa in the Pantheon catalogue, BAO measurements, binned data for Hubble diagram of QSOs, and the Hubble diagram of GRBs. \jt{ We mention that the complete sample of QSOs contains 1598 data points ranging $0.04 < z < 5.1$, while we used a binned catalogue including 25 data points from \cite{Risaliti:2018reu}}. More details of sample selection and the procedure of binned data have been discussed in \citealp{Risaliti:2018reu}. The data sample for GRBs used in this research contains 137 data points at the redshift range $0.03<z<5.5$ collected in \citep{Escamilla-Rivera:2021vyw}. We note that the GRBs sample in \citep{Escamilla-Rivera:2021vyw} has 141 data points extended to redshift $9.3$. However, we neglect 4 data points at redshifts higher than $z=5.5$, because our Pade cosmographic methods was not examined beyond $z \sim 5.5$, in the previous section. We use the BAO data from seven different surveys: 6dFGS, SDSS-LRG, BOSS-MGS, BOSS-LOWZ, WiggleZ, BOSS-CMASS, BOSS-DR 12 \citep[for details see][]{Camarena:2018nbr,Davari:2019tni}. Sample ({\it i}) includes the low-redshift data: SNIa+BAO+QSOs (up to z=1.8)+GRBs (up to z=2), which embraces redshift range $0.01<z<2.26$ and sample ({\it ii}) includes all observational data: SNIa+QSOs+GRBs that covers redshift range $0.01<z<5.5$. Notice that the binned QSOs up to $z=1.8$ contains 16 data points from all 25 data points. In addition, GRBs at $z<2$ includes 76 data points from all 141 data points. In the case of sample ({\it i}), we read the total chi-square from
\begin{eqnarray}\label{eq28}
\chi^{2}_{tot}= \chi^{2}_{SNIa}+\chi^{2}_{QSOs(z<1.8)}+\chi^{2}_{GRBs(z<2)}+\chi^{2}_{BAO},
\end{eqnarray}
and in the case of sample ({\it ii}), we use
\begin{eqnarray}\label{eq29}
\chi^{2}_{tot}= \chi^{2}_{SNIa}+\chi^{2}_{QSOs}+\chi^{2}_{GRBs}\;.
\end{eqnarray}
Our numerical results for minimizing $\chi^2$ function in the context of MCMC algorithm have been reported in Tabs.(\ref{tab:LCDM_real} , \ref{tab:pade_lowz} \& \ref{tab:pade_highz}). In Tab.(\ref{tab:LCDM_real}), we first report the best fit values of the free parameters of the flat-$\Lambda$CDM model for both dataset combinations and then show the best fit values of the cosmographic parameters of the model, utilizing Eq.(\ref{eq17}). In Tab.\ref{tab:pade_lowz} (Tab.\ref{tab:pade_highz}), we present the best-fit values of the cosmographic parameters in the context of Pade cosmographic methods, using sample (\it i}) (sample {\it ii}). In the following, we discuss our numerical results in this section.

%To compare the $\Lambda$CDM model in the context of the cosmography approach, firstly, we must have the cosmographic parameters of the model; therefore, best-fit values of the model's free parameters exhibited in eq.\ref{eq16} have been constrained using the MCMC algorithm and setting initial values. The results have been reported in table.\ref{113}. By utilizing these best fit values, Chi-square has been achieved using eq.\ref{eq26}. Then by fixing values of $\Omega_{b_0}$, $\Omega_{dm_0}$, $h$ in equations \ref{eq17}-\ref{eq21} for $\Lambda$CDM model the best-fit values of cosmographic parameters for model have been acquired reported in table.\ref{113} for low-z (Pantheon + Quasar + BAO) and high-z (Pantheon + Quasar + GRB).\\

\begin{table}
	\centering
	\caption{Same as Tab.(\ref{tab:pade_lowz}), but using sample ({\it ii}).}
	\begin{tabular}{ c c c c c c c c c c }
		\hline \hline
		Cosmographic parameters & Pade (2,2) & Pade (3,2) \\
		\hline
		$q_0$ & $-0.562\pm 0.029$ &$-0.727\pm 0.064$  \\
		\hline
		$j_0$ & $0.94^{+0.10}_{-0.14}$ &$1.63^{+0.35}_{-0.40}$  \\
		\hline
		$s_0$& $-0.02\pm 0.22$ &$-2.28^{+0.33}_{-0.73}$  \\
		\hline
		$l_0$& $3.81^{+0.16}_{-0.090}$ &$0.4^{+3.5}_{-4.2}$  \\
		\hline
		$m_0$   & $-$ &$-9.1^{+2.5}_{-5.4}$ \\
		\hline 
		Statistical criteria\\
		\hline
		$\chi^2_{min}$ & $1215.15$ &	$1207.95$  \\
		\hline
		AIC & $1223.15$	&	$1217.95$  \\
		\hline
		$\Delta AIC =AIC-AIC_{\Lambda CDM}$ & $1.03$ &	$-4.17$  \\
		\hline\hline
	\end{tabular}\label{tab:pade_highz}
\end{table}

\begin{figure} 
	\centering
	\includegraphics[width=6.5 cm]{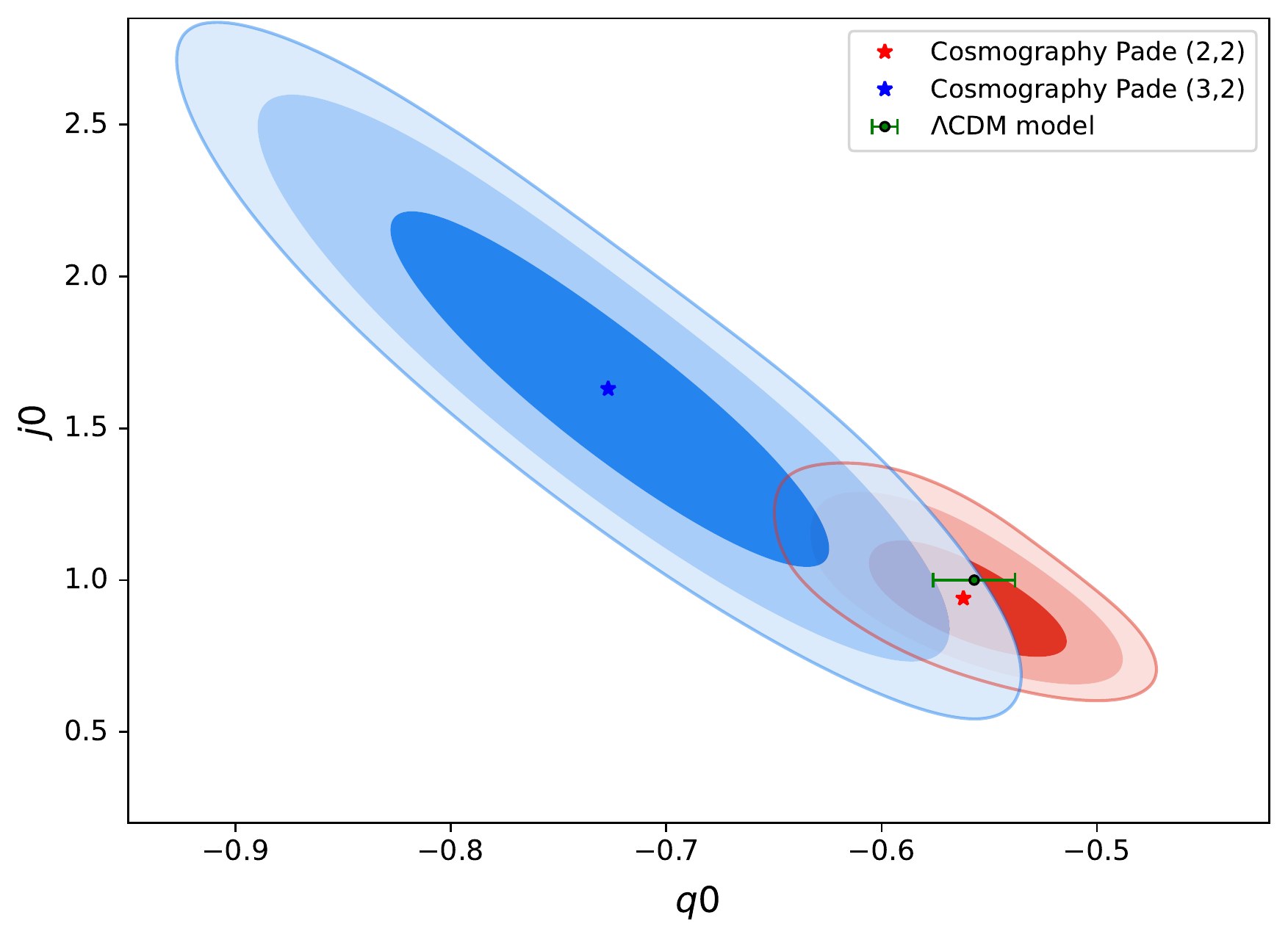}
	\caption{Same as Fig.(\ref{fig:real_low}), but for sample ({\it ii}).
	}
	\label{fig:real_high}
\end{figure}
\subsection{Numerical results for sample ({\it i})}
As it is shown in the upper part of Tab.(\ref{tab:LCDM_real}) for sample ({\it i}), using the low-redshift observations including SNIa, QSOs ($z<1.8$), GRBs ($z<2$) and BAO, we can put observational constraints on the cosmological parameters $\Omega_{m0}$ and $h$ of the standard flat-$\Lambda$CDM model as $\Omega_{m0}=0.3011^{+0.008}_{-0.008}$ and $h = 0.7167 \pm{0.0018}$. Our constraints on $h=H_0/100$ is between the Planck measurement $H_{0}=67.4\pm{0.5}$ $kms^{-1}Mpc^{-1}$ \citep{Planck:2018vyg} and local determinations $H_{0} = 73.0\pm{1.4}$ $kms^{-1}Mpc^{-1}$ \citep{Riess:2020fzl}. Using our constrains on $\Omega_{m0}$ and Eq.(\ref{eq17}), we obtain the best-fit values of the cosmographic parameters in flat-$\Lambda$CDM cosmology, as reported in the middle part of Tab.(\ref{tab:LCDM_real}) for sample ({\it i}). Finally, we report the minimum value of $\chi^2$ function and corresponding AIC value in the lower part of Tab.(\ref{tab:LCDM_real}). In the next step, we constraint the cosmographic parameters in the context of Pade cosmographic methods, using the datasets of sample ({\it i}) as shown in Tab.(\ref{tab:pade_lowz}) and Fig.(\ref{fig:real_low}). In Fig.(\ref{fig:real_low}). we have shown $1 \sigma$ to $3 \sigma$ confidence levels of the cosmographic parameters $q_0$ and $j_0$ in the context of Pade (2,2) and Pade (3,2) approaches. We observe that the $\Lambda$CDM value ($q_0=-0.548$, $j_0=1.0$) is inside the confidence regions of cosmographic methods, meaning that there is no observational tension between standard $\Lambda$CDM cosmology and datasets of sample ({\it i}). Quantitatively speaking, the differences between best-fit values of the cosmographic parameters of the flat-$\Lambda$CDM model and those of the Pade (2,2) cosmographic method are $1.4\sigma$ for $q_0$, $0.97\sigma$ for $j_0$, $1.18\sigma$ for $s_0$ and $4.4 \sigma$ for $l_0$. All these differences (except $l_0$) are the statistical error of the MCMC algorithm and cannot be considered as observational tension. Our comparison between standard $\Lambda$CDM model and Pade (3,2) cosmographic method shows $2.05\sigma$, $1.55 \sigma$, $1.22 \sigma$, $2.23\sigma$, and $1.3\sigma$ deviations for the cosmographic parameters $q_0$, $j_0$, $s_0$, $l_0$ and $m_0$, respectively. Like Pade (2,2), these deviations show that there is no significant tension between $\Lambda$CDM cosmology and Pade (3,2) method and consequently observational data in sample ({\it i}). In addition, we report the values of $\chi^2_{min}$, AIC and $\Delta AIC$ for Pade cosmographic approaches in Tab.(\ref{tab:pade_lowz}). Form the statistical AIC criteria, we observe no strong evidence ($|\Delta AIC|>6$) against the flat-$\Lambda$CDM model. This result is in agreement with cosmographic analysis for both Pade (2,2) and Pade (3,2) shown in Fig.(\ref{fig:real_low}). In similar study, Hu and Wang \cite{Hu:2022udt} have shown that the Pade approximations perform well compared to other approximations. We mention that their data sample consists of low- and high-redshift SNIa and GRBs and their Pade approximation directly used for luminosity distance.

\subsection{Numerical results for sample ({\it ii})}
Our constraint on the free parameter $\Omega_{m0}$ of the flat-$\Lambda$CDM model using the dataset of sample ({\it ii}) including the Hubble diagrams of SNIa, QSOs and GRBs represents $\Omega_{m0}=0.295 \pm{0.012}$. Using this constraint and  Eq.(\ref{eq17}), we report the best-fit values of the cosmographic parameters in the context of standard model in the last column of the middle part of Tab.(\ref{tab:LCDM_real}). The statistical results are reported in the last column of the lower part.
We observe that, in this case, $q_{0}$ is equal to $-0.557\pm{0.019}$, which is in $1 \sigma$ confidence level of its corresponding value obtained form sample ({\it i}). This means that the $q_0$ parameter in standard model does not change considerably due to varying the low-redshift Hubble diagrams to high-redshift data. This result is valid for higher cosmographic parameters $s_0$, $l_0$ and $m_0$ (see our constraints for samples ({\it i}) and sample ({\it ii}) in Tab. \ref{tab:LCDM_real}). In the context of Pade cosmographic methods, our constraints on the cosmographic parameters using sample ({\it ii}) are reported in Tab.(\ref{tab:pade_highz}). In addition, we show $1 \sigma$ to $3 \sigma$ confidence levels of $q_0$ and $j_0$ parameters in Fig.(\ref{fig:real_high}). We observe that the best-fit values of $q_0$ and $j_0$ parameters of Pade (2,2) cosmographic method are very close to those $\Lambda$CDM values. In fact, using sample ({\it ii}), the best-fit values of $q_0$ and $j_0$ in Pade (2,2) cosmographic method respectively decreases as $11 \%$ and $42 \%$ compare to corresponding values obtained from sample ({\it i}). On the other hand, we observe $2.55 \sigma$ and $1.68 \sigma$ deviations between best-fit values of $q_0$ and $j_0$ obtained in Pade (3,2) cosmographic approach with those of the standard $\Lambda$CDM model (see Fig.\ref{fig:real_high}). We mention that these deviations do not show any tension between Pade (3,2) and standard model and therefore can be assumed as statistical errors of the MCMC algorithm. Form the view point of AIC criteria, as reported in the lower part of Tab.(\ref{tab:pade_highz}), we observe no strong evidence ($|\Delta AIC| >6$) against the standard $\Lambda$CDM model.

\begin{table*}
	\caption{The best-fit values of the cosmographic parameters $q_0$ and $s_0$ in the context of Pade (3,2) method and their extracted $\Omega_{m0}$ values in standard flat-$\Lambda$CDM model obtained using high-redshift ($z>1$) and all data samples of SNIa, QSOs and SNIa+QSOs observations.}
	\centering
	\begin{tabular}{ c c c c c c c c c  }
		\hline \hline
		& High-z data& $\vert$ & $q_0$ & $\Omega_{m0}$($q_0$) & $\vert$ & $s_0$ & $\Omega_{m0}$$(s_0)$ \\
		\hline
		&SNIa & $\vert$ & $-0.24\pm 0.42$ & $0.51\pm 0.28$ & $\vert$ & $-1.48^{+0.98}_{-1.8}$ & $0.55^{+0.40}_{-0.22}$\\
		\hline 
		&QSOs & $\vert$ & $-0.095^{+0.49}_{-0.28}$ & $0.60^{+0.33}_{-0.18}$ &  $\vert$ & $-1.5^{+1.1}_{-1.5}$ & $0.55^{+0.34}_{-0.24}$ \\
		\hline 
		&SNIa + QSOs & $\vert$ & $-0.16^{+0.45}_{-0.34}$ & $0.56^{+0.30}_{-0.23}$ &  $\vert$ & $-1.5\pm 1.3$ & $0.55\pm 0.28$ \\
		\hline \hline
		& All data & $\vert$ & $q_0$ & $\Omega_{m0}$($q_0$) & $\vert$ & $s_0$ & $\Omega_{m0}$$(s_0)$ \\
		\hline
		&SNIa & $\vert$ & $-0.658\pm 0.076$ & $0.23\pm 0.05$ & $\vert$ & $-1.3^{+1.4}_{-1.5}$ & $0.51^{+0.33}_{-0.31}$\\
		\hline 
		&QSOs & $\vert$ & $-0.58^{+0.16}_{-0.39}$ & $0.28^{+0.11}_{-0.26}$ &  $\vert$ & $-1.5\pm 1.3$ & $0.55\pm 0.29$ \\
		\hline 
		& SNIa + QSOs & $\vert$ & $-0.664\pm 0.072$ & $0.22\pm 0.04$ &  $\vert$ & $-2.09^{+0.58}_{-1.3}$ & $0.69^{+0.28}_{-0.13}$ \\
		\hline \hline
	\end{tabular}\label{tab:Omega_0 evolution}
\end{table*}

\jt{
\subsection{Cosmographic method in redshift-bin}
Let us start with the results of the recent work \cite{Colgain:2022nlb} in which the authors showed that QSOs at redshifts lower than $z_{max}\sim 0.7$ recover the Plank-$\Lambda$CDM Universe (flat-$\Lambda$CDM model with $\Omega_{m0}\simeq 0.3$), in agreement with the predictions of the SNIa observations. By increasing the redshift range to $z>0.7$, one can get the transition from the Planck-$\Lambda$CDM Universe to the Einstein-de Sitter Universe (EdS) (spatially flat FLRW with only pressureless matter.)
From the observational point of view, the QSOs and SNIa are very different objects since SNIa in the Pantheon catalogue are more populated toward low-redshifts, while the QSOs in the Risaliti-Lusso sample \cite{Risaliti:2018reu,Risaliti:2015zla} are more numerous at higher redshifts. Interestingly, using the SNIa data at $z>1$, the authors of \cite{Colgain:2022nlb} have shown a substantial  increase of $\Omega_{m0}$ and a decrease in the $H_0$ measurement at redshifts higher than $z_{min}=1$. It is emphasized that any evolution of the $H_0$ value is equivalent to the evolution of the absolute magnitude of SNIa, which means that we have a stark choice between a SNIa cosmology and a flat-$\Lambda$CDM Universe. The same result with a different analysis was achieved by placing the standard flat-$\Lambda$CDM model in the redshift bin of observational data,  including observational Hubble data (OHD), SNIa and QSOs, supporting the idea of \cite{Colgain:2022rxy}. In the line of \cite{Colgain:2022nlb} and  \cite{Colgain:2022rxy}, we explore here the redshift evolution of $\Omega_{m0}$ using the Pade cosmographic method. For this purpose, we first remove the data of SNIa and QSOs at redshifts smaller than $z=1$, respectively, from the Pantheon catalogue and binned QSOs sample. We repeat our cosmographic analysis in the context of the Pade (3,2) approximation using the reduced binned sample of QSOs at redshifts $z>1$ (14 data points out of all 25 data) and 23 SNIa data points out of all 1048 data points in the Pantheon catalogue. The numerical results for the cosmographic parameters $q_0$ and $s_0$ obtained from the SNIa sample ($z>1$), binned QSOs sample ($z>1$) and combined SNIa+QSOs samples ($z>1$) are shown in Tab. (\ref{tab:Omega_0 evolution}). For comparison, we then show our results obtained from all data points of SNIa and QSOs. Substituting the numerical results of $q_0$ and $s_0$  into Eq. (\ref{eq17}), we obtain the best-fit value of $\Omega_{m0}$ for standard model, as shown in Tab. (\ref{tab:Omega_0 evolution}). Note that our cosmographic analysis for higher parameters $l_0$ and $m_0$ yields non-real solutions (complex values) of $\Omega_{m0}$ from the corresponding quadratic and cubic relations in Eq. (\ref{eq17}). Therefore, we restrict our analysis to the parameters $q_0$ and $s_0$ and their linear relations in Eq. (\ref{eq17}). The $\Omega_{m0}$ value resulted from $q_0-\Omega_{m0}$ and $s_0-\Omega_{m0}$ relations is approximately $\sim 0.5$ for the restricted data samples at $z>1$, as reported in the up-part of Tab. (\ref{tab:Omega_0 evolution}). This result is in contrast to the Planck-$\Lambda$CDM Universe and approximately in agreement with the predictions of \cite{Colgain:2022nlb,Colgain:2022rxy}. Moreover, the value of $\Omega_{m0}\sim 0.5$ predicted by our cosmographic approach supports the result of \cite{Khadka:2020vlh}, where it was shown that the higher value of $\Omega_{m0}$ is likely related to the high-redshift QSOs at $z \sim 2-5$.} We repeat the above analysis for data samples consisting of all SNIa, QSOs and SNIa+QSOs data points and report our results in the down-part of Tab. (\ref{tab:Omega_0 evolution}).
We see that the extracted $\Omega_{m0}$ value from $q_0$ is close to the Planck-LCDM value $\Omega_{m0}=0.3$, as expected. While there is a potential to get larger values of $\Omega_{m0}$ from the cosmographic parameter $s_0$.This result is due to this fact that $s_0$ parameter becomes more relevant than $q_0$ at high-redshifts. Notice that this result is obtained without restricting data samples to $z>1$. Hence, any larger value of $\Omega_{m0}$ extracted from $s_0$
directly relates to this fact that $s_0$ is more relevant at higher redshift, while $q_0$ is more significant at low redshifts.

\section{Conclusion} \label{sect:conlusion}
The cosmographic approach is used as a model-independent method to reconstruct the Hubble expansion of the universe. In this context, we use relevant mathematical series to reconstruct the Hubble parameter. The coefficients of this reconstruction which are related to the time derivative of the Hubble parameter (known as cosmographic parameters) can be constrained utilizing statistical methods and observational data. In other words, we can constrain the cosmographic parameters model-independently. Since the cosmographic method is a mathematical technique, it can serve as a benchmark for testing the physical models proposed in the field of cosmology.
In this work, we examined the standard flat-$\Lambda$CDM model from the viewpoint of cosmographic method using the Hubble diagrams of SNIa, QSOs, and GRBs, as well as the BAO observations. We investigated the Taylor series and rational Pade polynomials using mock data for SNIa and QSOs. We showed that in the redshifts $z<2.5$, where SNIa objects were observed \citep{Scolnic:2017caz}, the $4^{\it th}$ order Taylor expansion is insufficient to reconstruct the Hubble parameter, while the $5^{\it th}$ order Taylor series, Pade (2,2) and Pade (3,2) polynomials are acceptable approximations. At higher redshifts and in the redshift interval $0<z<5.5$, where QSOs have been observed \citep{Lusso:2017hgz}, we have shown that Pade cosmographic approaches are the useful approximations to reconstruct the Hubble function in a model-independent manner, while the $4^{\it th}$ order Taylor series is strongly rejected. In the next step, using the low-redshift combination of the Hubble diagrams from SNIa,  binned QSOs, GRBs and also the BAO measurements  as well as the high-redshift combination of the Hubble diagrams for SNIa, binned QSOs and GRBs, we showed that the cosmographic parameters of the standard flat-$\Lambda$CDM model are well consistent with those of the Pade cosmographic methods. In other words, our results for both low- and high-redshift Hubble diagrams, do not show any cosmological tension between the constrained values of the cosmographic parameters in the flat-$\Lambda$CDM model and those of the cosmographic method based on the rational Pade polynomials. This result is against the cosmographic tension presented for standard model, using the high-redshift Hubble diagrams from QSOs and GRBs in \cite{Lusso:2019akb,Risaliti:2018reu}. In addition, from the viewpoint of statistical AIC criteria, our Pade cosmographic analysis showed no strong evidence against the standard $\Lambda$CDM cosmology. 
Finally, we put the cosmographic method in redshift-bin of Hubble diagram data in order to explore the possibility of the variation of $\Omega_{m0}$ extracted at high-redshift observations.	Using the restricted data points of SNIa, QSOs and SNIa+QSOs samples at $z>1$, we found larger values of $\Omega_{m0}$ for standard flat-$\Lambda$CDM model extracted from both $q_0$ and $s_0$ parameters. This result confirms a transition form Planck-$\Lambda$CDM value $\Omega_{m0}=0.3$ at low-redshift to larger values at high-redshifts \cite[see also][]{Colgain:2022nlb,Colgain:2022rxy}. In addition, using all data points of SNIa, QSOs and SNIa+QSOs samples, we extracted a larger value of $\Omega_{m0}$ from $s_0$ parameter, while the extracted value from $q_0$ parameter is sufficiently close to Planck-$\Lambda$CDM value. This result is expected because the $s_0$ ($q_0$) parameter is more relevant at high-(low) redshifts.

 \bibliographystyle{apsrev4-1}
\bibliography{ref}

\label{lastpage}

\end{document}